\documentclass[a4paper]{article}
\usepackage[T1]{fontenc}
\usepackage[utf8]{inputenc}
\usepackage[english]{babel}

\usepackage{geometry}
\usepackage[inline]{enumitem}

\usepackage{graphicx}
\usepackage{caption}
\usepackage{subfig}

\usepackage{amsmath}
\usepackage{amssymb}
\usepackage{bm}
\usepackage{mathtools}
\DeclarePairedDelimiter{\abs}{\lvert}{\rvert}

\usepackage{blkarray} 

\usepackage{xcolor} 
\definecolor{col1}{HTML}{E41A1C}
\definecolor{col2}{HTML}{377EB8}
\definecolor{col3}{HTML}{4DAF4A}
\definecolor{col4}{HTML}{984EA3}
\definecolor{col5}{HTML}{FF7F00}
\definecolor{col6}{HTML}{FFFF33}

\usepackage{booktabs}
\usepackage{multirow}
\usepackage{siunitx}
\usepackage{rotating}

\usepackage[autostyle,italian=guillemets]{csquotes}
\usepackage[backend=biber,style=apa,maxcitenames=1]{biblatex}
\AtBeginBibliography{\small} 

\usepackage[ruled,vlined]{algorithm2e}
\usepackage{listings}
\usepackage{hyperref} 

\title{Multivariate hierarchical analysis of car crashes data considering a spatial network lattice}
\author{Andrea Gilardi, Jorge Mateu, Riccardo Borgoni, Robin Lovelace}

\providecommand{\keywords}[1]{ \small\textit{Keywords: } #1}
\date{}

\begin{document}

\maketitle

\begin{abstract}
Road traffic casualties represent a hidden global epidemic, demanding evidence-based interventions. 
This paper demonstrates a network lattice approach for identifying road segments of particular concern, based on a case study of a major city (Leeds, UK), in which 5,862 crashes of different severities were recorded over an eight-year period (2011-2018).
We consider a family of Bayesian hierarchical models that include spatially structured and unstructured random effects, to capture the dependencies between the severity levels. 
Results highlight roads that are more prone to collisions, relative to estimated traffic volumes, in the north-west, and south of city-centre. 
We analyse the Modifiable Areal Unit Problem (MAUP), proposing a novel procedure to investigate the presence of MAUP on a network lattice.
We conclude that our methods enable a reliable estimation of road safety levels to help identify `hotspots' on the road network and to inform effective local interventions.
\end{abstract}


\section{Introduction}

Road casualties have been described as a global epidemic, representing the leading cause of death among young people worldwide (see, e.g. \textcite{mackay_traffic_1972, nantulya_neglected_2002}).
Car crashes and other types of collisions are responsible for more than 1 million deaths each year (1,250,000 in 2015, 17 deaths per 100,000 people), as reported by \textcite{worldhealthorganization_global_2018}.
In high income nations, such as the United Kingdom (UK), the roads are safer than the global average, but car crashes are still the cause of untold suffering. 
According to the statistics published by the UK's Department for Transport (DfT) in the \textit{Annual report on Road Casualties in Great Britain} \parencite{dft_report}, approximately 153,000 road traffic collisions resulting in casualties were recorded in 2019, 5\% lower than 2018 and the lowest level since records began, in 1979.
Nevertheless, the DfT estimates that approximately 33,648 people were killed or seriously injured (KSI) in 2019, and while this number is slightly lower than in 2018, the decaying rate has been getting lower and lower starting from 2010. 
These figures are worrisome considering that car occupant fatality rates are particularly high in the 17-24 age band \parencite[p. 17]{dft_report}. 

To tackle the flattening trend in the KSI rate over the past decades, a range of interventions are needed, and analytical approaches can help prioritise them.
This paper presents a statistical model to identify street sections with anomalously high car crashes rates, and to support police responses and cost-effective investment in traffic calming measures \parencite{pacts_roads_2020}.

Statistical models of road crashes have become more advanced over time.
In the 1990s, models tended to consider only the discrete and heterogeneous nature of the data \parencite{miaou1993modeling, miaou1994relationship, shankar1995effect}, omitting spatial characteristics.
More recent statistical models of crash data include consideration of crash location in two-dimensional space, with three main advantages for road safety research \parencite{el2009urban}.
First, consideration of space allows estimating appropriate measures of risk (such as expected counts, rates or probabilities) at different levels of resolution and the subsequent ranking of geographical areas to support local interventions.  
Second, spatial dependence can be a surrogate for unknown, potentially unmeasured (or unmeasurable) covariates; adjusting for geographic location can reduce model misspecification \parencite{dubin1988estimation, cressie1993statistics}. 
Third, the spatial dimension can be used to take advantage of autocorrelation in the relevant variables, borrowing strength from neighbouring sites and improving model parameter estimation.

Road crash datasets, which are typically available on a single accident basis, can be spatially aggregated in two main ways: administrative zones (such as cantons, census wards, or regions) or street network features (either as contiguous segments or divided into corridors and intersections).
In both cases, the spatial support is a lattice, i.e. a countable collection of geometrical units (polygons or lines, respectively), possibly supplemented by a neighbourhood structure. 
Several papers addressed the statistical modelling of crash frequencies at the areal level, based on available zoning systems in the study region (see, e.g., \textcite{miaou2003roadway, aguero2006spatial, noland2004spatially, boulieri2017space}).
The second approach has gained in popularity in recent years, with a number of papers analysing road crash events aggregated to the street level (see, e.g., \textcite{miaou2005bayesian, aguero2008analysis, wang2009impact}), as detailed in several review papers \parencite{lord2010statistical, savolainen2011statistical, ziakopoulos2020review}.
In reference to street level data, we note that there has been a recent surge of research for spatial point patterns living on networks \parencite{rakshit2019fast, cronie2020inhomogeneous, baddeley2020analysing}. 

Both zone and network level approaches have advantages, notably computational requirements for the former and spatially disaggregated results for the latter.
Given that computational resources are less of a constraint in the 2020s than they were in previous decades, and the fact that it is the nature of roads (not zones) that is responsible for crashes, we argue that road segments are the more appropriate aggregation units for the analysis of road crash data.
Network analysis can be used to bring attention to specific segments, and, for these reasons, the models presented in the next sections were developed considering a network lattice.

Aggregation, e.g. number of crashes per road segment, enables comparison between different road segments.
However, spatial aggregation also leads to a well-known problem in geographical analysis, the Modifiable Areal Unit Problem (MAUP), firstly described in  \textcite{openshaw1981modifiable}: the size of the spatial units impacts on the statistical analysis, influencing, and possibly biasing, modelling choices and results. 
Hence, conclusions drawn at one scale of spatial aggregation might not necessarily hold at another scale or be somehow different.
The MAUP has been mainly ignored in the road safety literature and, as reported by \textcite{xu2018modifiable} and \textcite[p. 21]{ziakopoulos2020review}, it is mentioned only in a handful of recent papers \parencite{ukkusuri2012role, abdel2013geographical, zhai2019influence, briz2019investigation}, which explore the impact of changing the areal zoning system (e.g. TAZ, block groups and census tracts) on parameter estimates, significance and hotspot detection.
Only one early paper \parencite{thomas1996spatial} could be found exploring the impacts of the MAUP on road crash data, albeit only in terms of  summary statistics of aggregated counts. 
To assess the MAUP effect on network data modelling, we employed an algorithm to modify the structure of a road network, merging contiguous segments in the same corridor preserving the geometrical properties of the network \parencite{R-dodgr}. 
Then, we compared the results obtained with the two different network configurations. 
To the best of our knowledge, this is the first attempt at exploring and estimating the presence and the magnitude of MAUP in models that consider a network lattice. 

Finally, we note that systems of collision classification present a multivariate nature \parencite{kirk_implications_2020}. 
The occurrences of different severity degrees can be correlated to each other, and their spatial dynamics can be potentially interdependent.
Hence, it is necessary to account for correlations between crashes counts at different levels of severity. 
We consider two types of accidents: \textit{slight} and \textit{severe}. 
The severe class is very sparse in the dataset at hand, hence modelling both types of accidents simultaneously allows to borrow strength from the existing correlations and improves estimates. 
We underline that the methodology for classifying the severity level of a car crash in the UK has been modified starting from 2016, adopting the injury-based systems called \texttt{CRASH} and \texttt{COPA} \parencite{CRASH}. 
All police forces are gradually adopting these new reporting systems in England, and the Office for National Statistics (ONS) developed a logistic regression model to correct the severity levels between different years and classification systems. 
The data used in this paper have been adjusted using the procedures developed by ONS \parencite[p. 38-41]{dft_report}.

Following ideas introduced in \textcite{barua2014full}, we consider a range of competing models, developed in a full hierarchical Bayesian paradigm. 
This approach allows one to encompass complex structures of spatial dependence in a quite natural way. 
Spatially structured random effects are defined using both Intrinsic Multivariate Conditional Auto-regressive (IMCAR) and Proper Multivariate Conditional Auto-regressive (PMCAR) priors \parencite{besag1974spatial, mardia1988multi, martinez2019disease, palmiperales2019bayesian}.  
\textcite{lord2010statistical, savolainen2011statistical, ziakopoulos2020review} review other examples of multivariate car crashes models on a network.

The case study is the metropolitan area of Leeds (population 800,000) in North England.
We accessed Ordnance Survey data on major roads (3661 segments, total length 450 km), creating a spatial network substantially larger than previous studies, many of which report findings on only a few roads, with \textcite{borgoni2020assessing} representing a notable exception albeit with a simpler spatial structure. 
We present results for an entire metropolitan area, approximating more closely the level at which road policing activities and investment in road safety interventions are prioritised.
The scale of the case study presented several computational challenges, and, in terms of Bayesian parameter estimation, we used the computationally efficient Integrated Nested Laplace Approximation (INLA) approach instead of Markov chain Monte Carlo (MCMC) sampling \parencite{INLA1, INLA2}. 

The rest of the paper is organised as follows. 
In Section~\ref{sec:data}, the data sources are described.
In Section~\ref{sec:methodology}, the statistical methodology adopted in this paper is discussed in detail. 
In Section~\ref{sec:results}, the main results of the paper are presented whereas, model criticism and further model discussion, such as MAUP analysis, are provided in Section~\ref{sec:criticism}. 
Conclusions, in Section~\ref{sec:conclusions}, end the paper. 

\section{Data}
\label{sec:data}

\begin{figure}
	\centering
	\includegraphics[width=\linewidth]{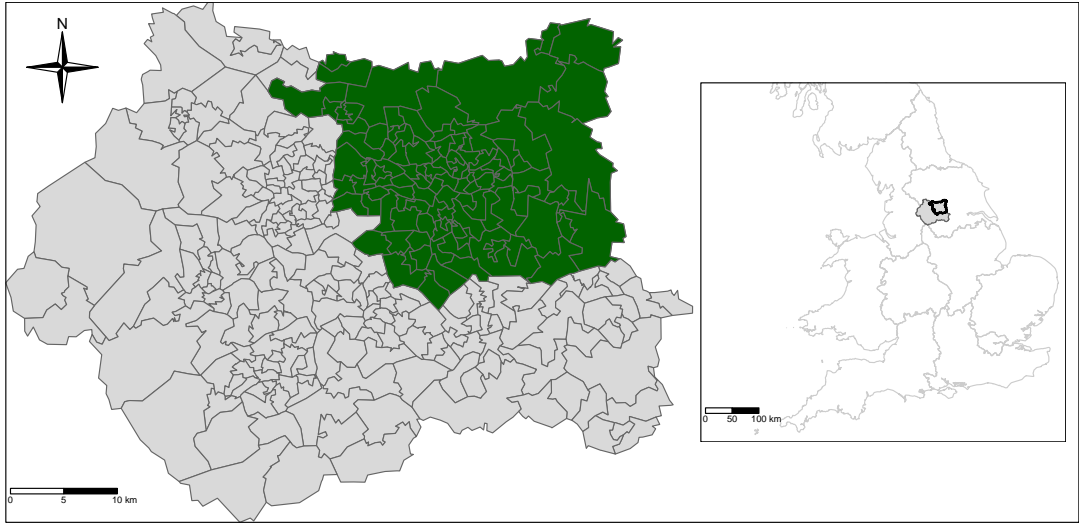}
	\caption{The grey polygons show the MSOAs in West-Yorkshire region, while the dark-green area highlights the city of Leeds. The inset map locates the position of the study-area with respect to England.}
	\label{fig:msoa}
\end{figure}

The datasets analysed in this paper came from several different sources and required a number of preprocessing steps before they could be made into a structure suitable for a statistical analysis.
The study region was defined as the Middle Super Output Area (MSOA) zones within the local authority of Leeds. 
The City of Leeds was selected because it is a car-dependent city with a large network of major roads that approach the city centre (the city was dubbed the 'motorway city of the 70s') and would therefore be expected to be a place where road safety could be improved.
Leeds is part of West Yorkshire and accounts for approximately 40\% of all car crashes in the region. 

Origin-destination data from the 2011 UK Census were used to estimate traffic volumes, to provide an estimate of exposure, with traffic volumes used as part of the denominator of the statistical models presented in Section~\ref{sec:methodology}. 
The road network was obtained from Ordnance Survey, covering all major roads in Leeds. 
We matched the network and the MSOAs using an overlay operation. 
We associated all car crashes that occurred in the city of Leeds from 2011 to 2018 with the nearest point on the road network, counting the occurrences in each street segment. 
Finally, a set of socio-economic variables obtained from 2011 UK Census data were included in the statistical models as fixed effects.

\subsubsection*{MSOA zones}

There are 6,791 MSOAs in England, 299 of which belong to the West Yorkshire region and 107 of which constitute Leeds.
These were accessed from the github-page\footnote{URL: \url{https://github.com/npct/pct-outputs-regional-R/tree/master/commute}, last access on 06/2020} of Propensity Cycle Tool \parencite{lovelace_propensity_2017}.
The MSOAs represent the starting point for all the following steps, and they are mapped in Figure~\ref{fig:msoa} as grey polygons for the West-Yorkshire, and as dark-green polygons for the City of Leeds. 
The inset map is used to locate the study-area in the British territory.

\subsubsection*{Traffic flow}

\begin{figure}
	\centering
	\subfloat[Random sample of one thousand Origin-Destination data representing the daily flows between MSOAs in Leeds. \label{fig:raw_flows}]{
		\includegraphics[width=0.47\linewidth]{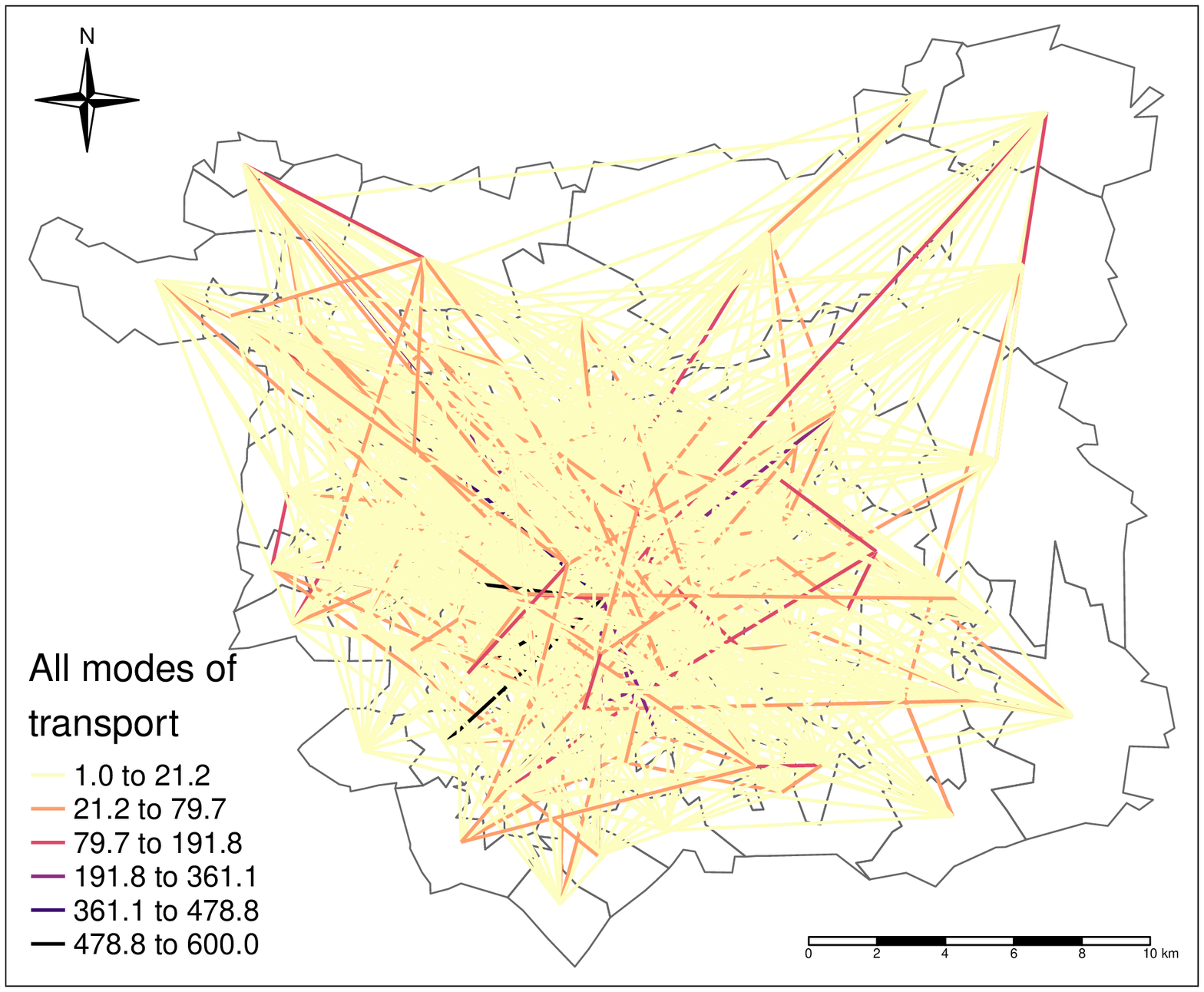}
	}
	\hspace{0.01\linewidth}
	\subfloat[Estimates of the new traffic measure. The white star represents Leeds City Centre. \label{fig:new_flows}]{
		\includegraphics[width=0.47\linewidth]{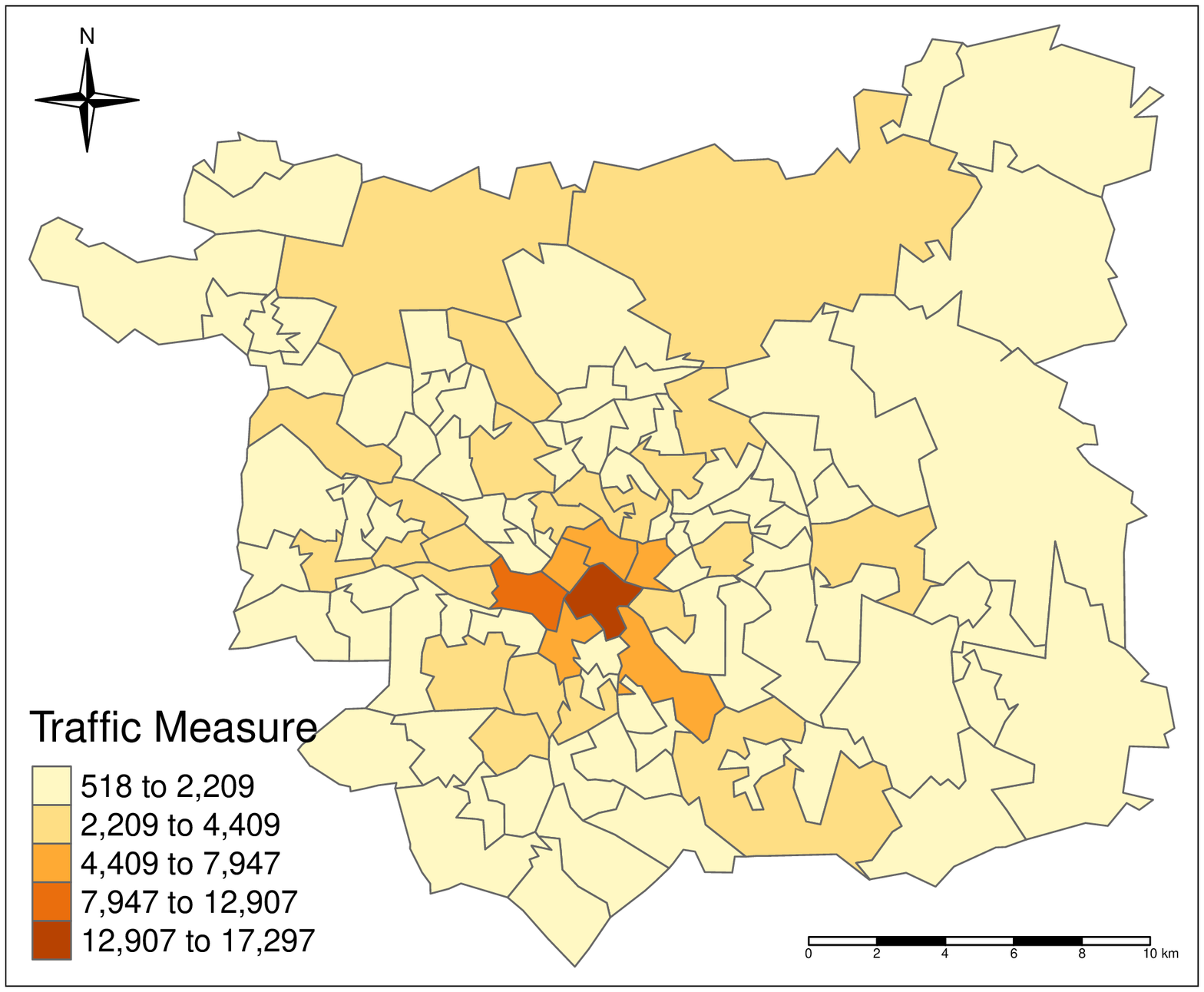}
	}
	\caption{Raw and modified traffic flows in the area of Leeds. The map on the right highlights several contiguous MSOAs that correspond to the arterial thoroughfares that are used to reach the City Centre.}
	\label{fig:flows}
\end{figure}

The \textit{traffic flow} data represent the \textit{commuting journeys} from home to workplace using several modes of transport, such as train, bus, bike and motorcycle. 
The data were collected during the 2011 Census at the individual level, and then aggregated at the MSOA level. 
The UK Data Service shares the flow data through the \texttt{WICID} interface as cross-tables reporting the flows between all pairs of a predefined set of MSOAs \parencite{WICID}. 
We considered the commuting flows in the region of Leeds for all possible modes of transport. 
Figure~\ref{fig:raw_flows} shows a random sample of $1000$ \textit{traffic flows} (out of 10,536 in total) between the centroids of the MSOAs in Leeds, coloured according to the number of daily commuters.

Raw WICID data, however, ignore that people may travel to their workplace through several MSOAs. 
For this reason, we calculated a new traffic measure using the following procedure. 
Starting from the MSOAs, we defined a graph where the vertices are the centroids of each area, and the edges connect neighbouring areas.
Then, we estimated the shortest path for all commuting journeys downloaded from WICID and assigned to each MSOA a value that is equal to the number of all raw traffic measures going through the area. 
These values represent the new traffic measures and are displayed in Figure~\ref{fig:new_flows}. 
A similar approach was also adopted in \textcite{boulieri2017space}, and we refer to the references therein for more details. 
The raw data flows and the MSOAs polygons were downloaded using the R package \texttt{pct} \parencite{pct}.

\subsubsection*{Road network}

The \textit{road network} was built using data downloaded from Ordnance Survey (OS), an agency that provides digital maps and other services for location-based products \parencite{OS}. 
We downloaded the \texttt{Vector OpenMap Local} data for a geographical region covering Leeds, selected the \texttt{Roads} and \texttt{Tunnels} layers, and filtered the streets that belong to the city. 
Ordnance Survey represents all the streets of a road network as the union of a finite set of segments, and it includes additional fields such as the road name or the street type. 
These segments represent the elementary units for the statistical analysis described in Section~\ref{sec:methodology}. 

The road network downloaded from OS is composed of approximately 50,000 segments. 
The OS network was simplified using the following procedure. 
We first selected only the major roads, such as the \textit{Motorways}, \textit{Primary Roads} and \textit{A roads}.
They represent 3,668 segments, i.e. less than 10\% of the total road network, but more than 50\% of all car crashes registered during 2011-2018 occurred in their proximity. 
The output of this procedure is a road network composed by a big cluster of connected streets, displayed in Figure~\ref{fig:show_network_crashes}, and several isolated segments of small groups of road segments (which are also called \textit{islands}), created by the exclusion of their links to the other roads.

\begin{figure}
\begin{minipage}[c]{0.35\textwidth} 
\centering
\includegraphics[scale=.35]{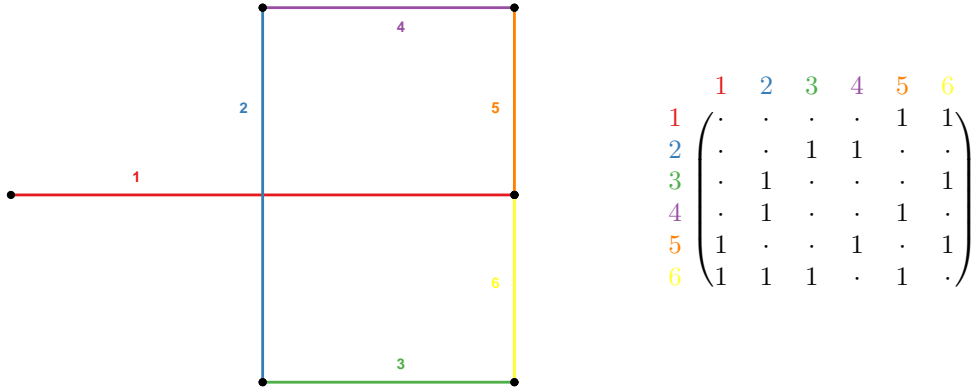} 	\end{minipage}
\hspace{0.1\textwidth}
\begin{minipage}[c]{.55\textwidth}
  \[
	\begin{blockarray}{ccccccc}
	& \textcolor{col1}{1} 
	& \textcolor{col2}{2} 
	& \textcolor{col3}{3} 
	& \textcolor{col4}{4} 
	& \textcolor{col5}{5} 
	& \textcolor{col6}{6} \\
	\begin{block}{c(cccccc)}
	\textcolor{col1}{1} 
	& \cdot & \cdot & \cdot & \cdot & 1 & 1 \\
	\textcolor{col2}{2} & \cdot & \cdot & 1 & 1 & \cdot & \cdot \\
	\textcolor{col3}{3} & \cdot & 1 & \cdot & \cdot & \cdot & 1 \\
	\textcolor{col4}{4} & \cdot & 1 & \cdot & \cdot & 1 & \cdot \\
	\textcolor{col5}{5} & 1 & \cdot & \cdot & 1 & \cdot & 1 \\
	\textcolor{col6}{6} & 1 & 1 & 1 & \cdot & 1 & \cdot \\
	\end{block}
	\end{blockarray}
  \]	
\end{minipage}
\caption{Graphical example showing the dual nature of a road network. Left: map showing the geographical dimension. Each segment is coloured and labelled using a different ID and colour respectively. Right: adjacency matrix of the graph associated with the road network. Each vertex corresponds to a segment whereas an edge connects two vertices if they share one boundary point. For example, segments $1$ and $2$ are not neighbours since they do not share any point at the boundaries, even if they intersect each other. This situation may occur at bridges or overpasses.}
\label{fig:W_matrix}
\end{figure}

These small clusters can be problematic from a modelling perspective since they produce a not-fully-connected network (see \textcite{hodges2003precision, FRENISTERRANTINO201825} and the properties of ICAR and MICAR distributions explained in Section~\ref{sec:methodology}), so we implemented an algorithm to further simplify the road network and remove them. 
This algorithm is based on the dual representation of a road network as a geographical entity, composed by points and lines, and a graph object, with nodes and edges \parencite{porta2006network, marshall2018street,  gilardi_lovelace_padgham_2020}. 
More precisely, we created a graph whose vertices correspond to the street segments of the road network, and we defined an edge for each pair of spatial units sharing a point at their boundaries. 
This graph uniquely determines a (sparse) adjacency matrix amongst the spatial units (i.e. the road segments) that summarises the graph dimension of the road network.
We sketched a toy example in Figure~\ref{fig:W_matrix}, representing the idea behind the dual representation of a road network and the definition of the adjacency matrix. 

Using the graph and the adjacency matrix, we excluded all road segments that did not belong to the main cluster. 
In particular, we dropped 7 segments spread across different parts of the city with a total length of approximately 380m (out of 450km). 
That also implied that we removed from the analysis 3 car crashes that occurred in those segments (see below). 
However, given the extremely small fraction of records discarded from the dataset, we expect that excluding these observations does not influence the final results. 
It should be stressed that this procedure creates a fully connected network, which has relevant benefits on the rank-deficiency problem of the ICAR models described in Section~\ref{sec:methodology}.

In the end, the road network is composed of 3661 units, and it is shown in Figure~\ref{fig:show_network_crashes}, where the segments are coloured according to their road types. 
The segments have different lengths, ranging from 0.1m to 2597m, with an average value of 118m (\textit{sd} = 178m). As explained in the next Section, these lengths are used in the exposure parameter of the statistical models in order to guarantee a comparable rate amongst the network units.
Finally, since the street network and the MSOAs are spatially misaligned, they were matched using an overlay operation: each road segment was assigned to the MSOA that intersects the largest fraction of the segment. 
This procedure allows us to assign a traffic estimate to each road segment, which will be used in the exposure parameter (along with the segments' lengths) in the statistical models considered below. 

\subsubsection*{Road traffic collision data}

\begin{figure}
	\centering
	\subfloat[Road network and car crashes in Leeds. \label{fig:show_network_crashes}]{
		\includegraphics[width=0.47\linewidth]{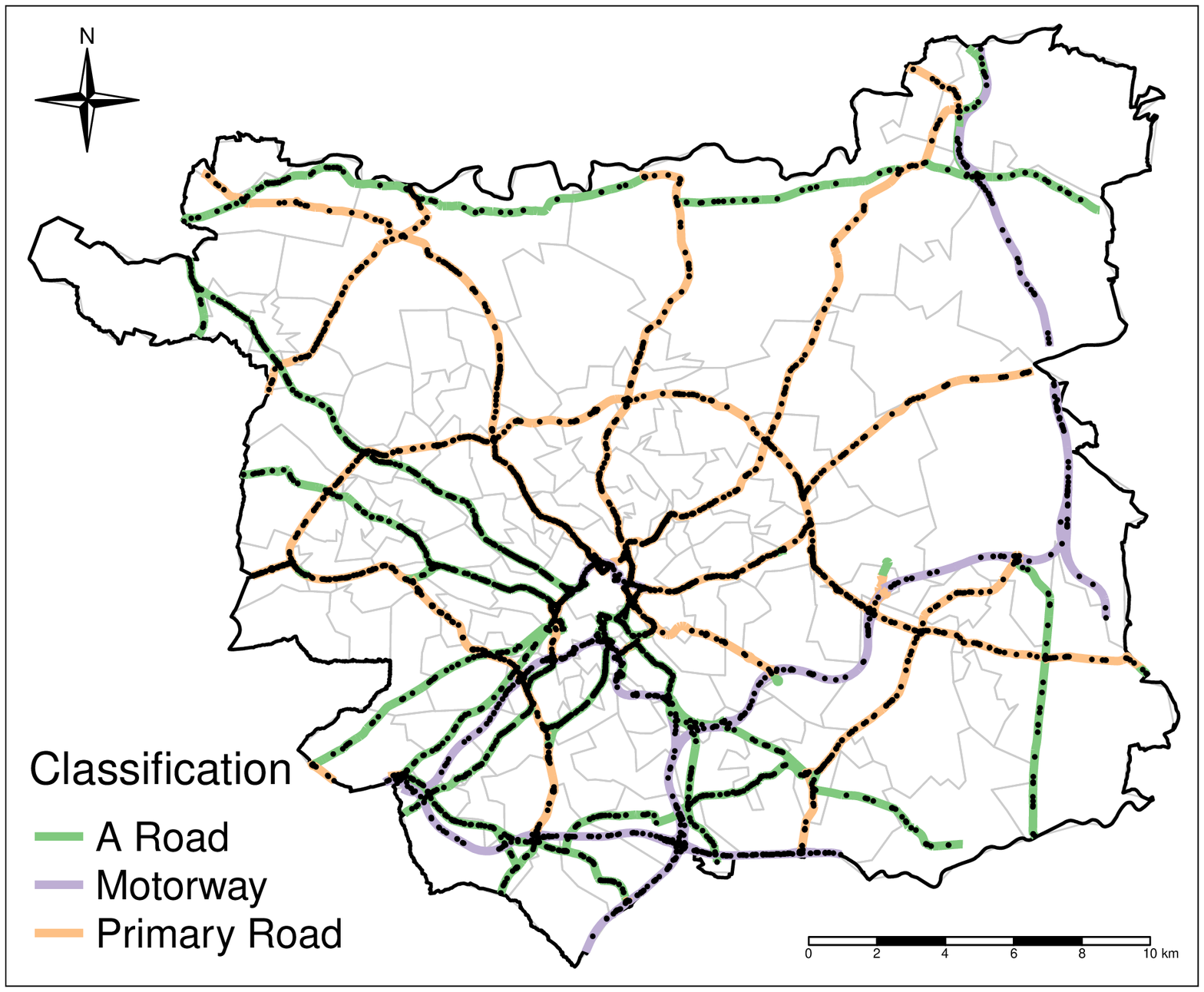}
	}
	\hspace{0.01\linewidth}
	\subfloat[Choropleth map of severe counts.  \label{fig:show_serious_counts}]{
		\includegraphics[width=0.47\linewidth]{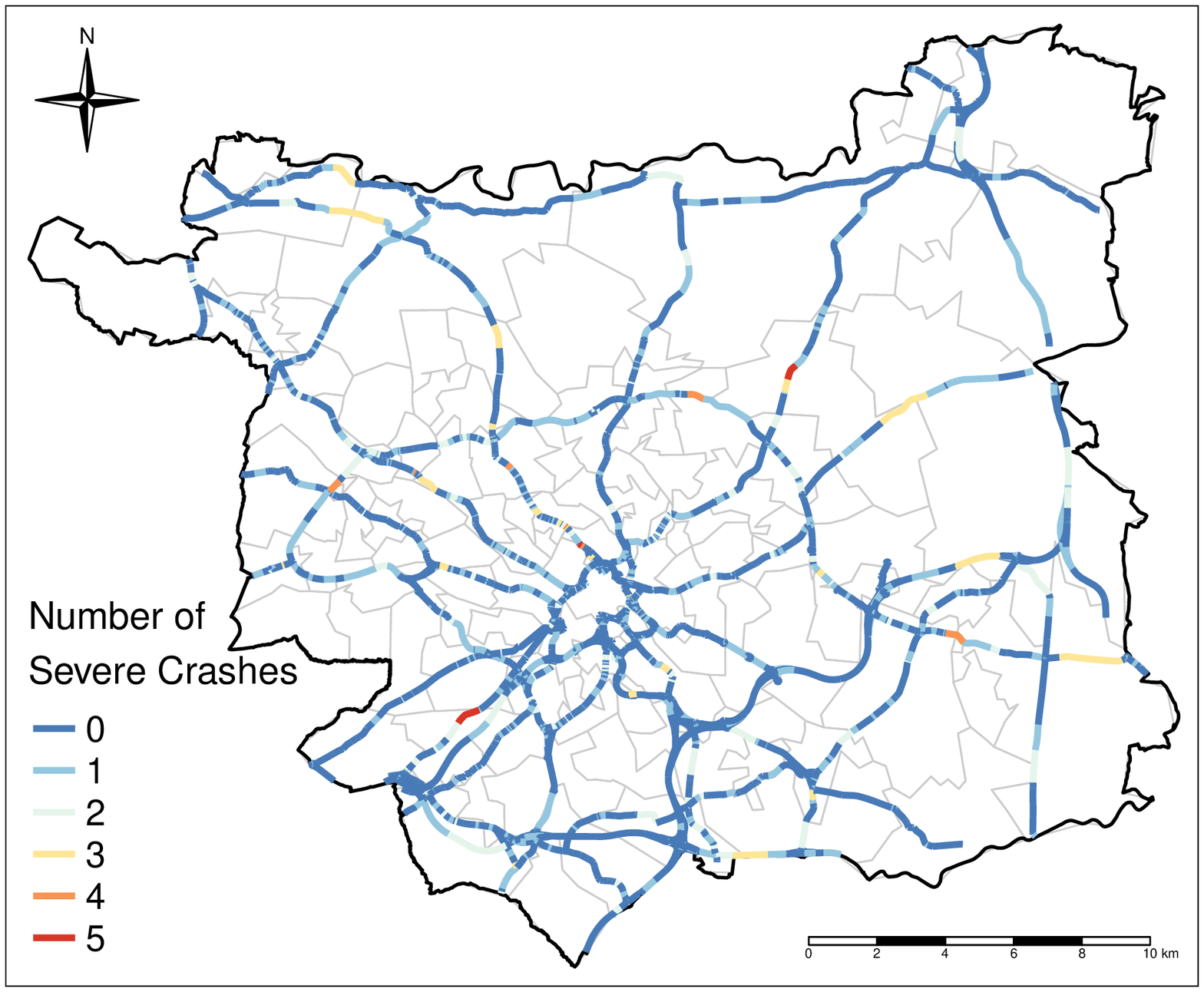}
	}
	\caption{The map on the left represents the road network in Leeds. Each segment is coloured according to its OS classification. The black dots represent the car crashes. On the right, we report a choropleth map displaying severe car crashes counts.}
	\label{fig:network_crashes}
\end{figure}

We analysed all road traffic collisions that occurred between the 1st of January 2011 and the 31st of December 2018 in the MSOAs pertaining to the City of Leeds, which involved at least one car, personal injuries, occurred on public roads and became known to the Police forces within thirty days of the occurrence. 
The geographical coordinates of the crashes are provided at ~10m  or less resolution in the UK’s official coordinate reference system (the Ordnance Survey National Grid, EPSG code 27700) depending upon the year to which the data refer to \textcite{Dep4Trans2011}. Hence, we adopted this value as a threshold to account for the potential misalignment between the event locations and the network and excluded all those events that occurred farther than ten meters from the closest segment in the simplified road network since they might be related to other streets.
The data were downloaded from the UK's official road traffic casualty database, called \textit{STATS19}, using the homonym R package \parencite{stats19}. 

As mentioned above, we focussed on car crashes that occurred in a public highway and involved personal injuries. 
In general, however, there is no obligation for people to report all personal injury accidents to the police and, for this reason, a proportion of non-fatal and no-injury casualties remain unknown to the police \parencite{dft_report}. 
Hence, we acknowledge that the counts of slight accidents considered hereinafter may suffer under-reporting to some extent.
We refer to \textcite[Sec 2.1]{savolainen2011statistical} and references therein for more details.

The final sample is composed of 5,862 events, and they are reported as black dots in Figure~\ref{fig:show_network_crashes}. 
Then, we projected all crashes to the nearest point on the road network, and we counted the number of \textit{slight} or \textit{severe} occurrences on all street segments. 
We decided to ignore the temporal dimension since \textit{severe} crashes counts present an extreme sparsity, with more than 80\% of zero counts during 2011-2018. 
Moreover, 40\% of all segments registered no car crashes during the study period, while another 40\% reported two or more crashes. 
These numbers highlight a common temporal trend between the eight years, and we refer the interested reader to the supplementary material for a space-time representation. 
The map in Figure~\ref{fig:show_serious_counts} shows the spatial distribution of \textit{severe} crashes counts.

\subsubsection*{Socio-economic covariates}
The statistical models described in Sections~\ref{sec:methodology} and~\ref{sec:results} include two socio-economic covariates that were obtained from the 2011 UK Census data and downloaded from Nomis website\footnote{URL: \url{https://www.nomisweb.co.uk/}. Last access: 06/2021.}. 
In particular, we consider the population density (given by the ratio of the number of inhabitants in a given region and its area in squared metres) and the employment rate (given by the ratio of employed population between 16 and 64 and total population between 16 and 64). 
Both covariates were obtained at the LSOA level (a more detailed aggregation level than MSOA) and, for the same reasoning as before, they were matched with the street segments using an overlay operation. 

\section{Statistical methodology}
\label{sec:methodology}

We first focus on the definition of a three-level hierarchical model structure which is shared among all the alternative specifications considered below. 
Then, we introduce two baseline models that serve as benchmarks and starting points for the other specifications. 
Thereafter, four different extensions to the baseline models are introduced.
Finally, some techniques used for model comparison are discussed. 
The common theme behind all the seven alternatives is the presence of spatially structured and unstructured multivariate random effects.

Let $Y_{ij}, \ i = 1, \dots, n,\ $ represent the number of car crashes that occurred in the $i$-th road segment with severity level $j, \ j = 1, \dots, J$. 
In this paper we consider two possible severity levels, a car crash being either \textit{severe}, $j = 1$, or \textit{slight}, $j = 2$.

In the first stage of the hierarchy, we assume that  
\[
Y_{ij} | \lambda_{ij} \sim \text{Poisson} \left(E_i\lambda_{ij}\right),
\]
where $E_i$ is an exposure parameter and $\lambda_{ij}$ represents the car crashes rate in the $i$th road segment for severity level $j$. 
As mentioned in the previous section, the exposure parameter, $E_i$, is given by the product of two quantities \parencite{wang2009impact}: the segment's length and the estimate of traffic flow (see Section~\ref{sec:data}).  
The exposure accounts for the fact that a longer street segment has a higher collision risk than a shorter one, guaranteeing that comparable rates amongst units (segments) are preserved. 
At the same time, the traffic flow estimates allow different segments of the network to be “weighted” differently, being more exposed to accidents those segments with a higher traffic flow, all the rest being fixed.

At the second stage of the hierarchical model, a log-linear structure on $\lambda_{ij}$ is specified. 
We assume that
\[
\log\left(\lambda_{ij}\right) = \beta_{0j} + \sum_{m = 1}^{M}\beta_{mj}X_{ijm} + \theta_{ij} + \phi_{ij},
\]
where $\beta_{0j}$ represents a severity-specific intercept, $\left\lbrace\beta_{mj}\right\rbrace_{m = 1}^{M}$ is a set of coefficients, $(X_{ij1}, \dots, X_{ijM})$ is a collection of $M$ covariates, $\phi_{ij}$ is a spatially structured random effect and $\theta_{ij}$ represents a normally distributed error component. 
The third stage that completes the hierarchical model is the specification of prior and hyperprior distributions. We assigned a vague $N(0, 1000)$ prior to $\beta_{mj}, \ m = 0, \dots M$. 
The two random effects, namely $\theta_{ij}$ and $\phi_{ij}$, represent the unstructured and structured spatial components and are defined differently in different models as discussed below. 
Hereafter, we follow the notation used in \textcite{martinez2019disease}. 

\subsection{Baseline models: independent spatial and unstructured effects}

The two baseline models are defined considering multivariate spatial and unstructured random effects with independent components. 
More precisely, a bivariate Gaussian prior with independent components is assigned to $\left(\theta_{i1}, \theta_{i2}\right)$ for both baseline models: 
\begin{equation}
\left(\theta_{i1}, \theta_{i2}\right)\sim N_2\left(\bm{0}, \begin{bmatrix}
\sigma^{2}_{\theta_{1}} & 0 \\
0 & \sigma^2_{\theta_{2}}
\end{bmatrix}\right), \ \quad i = 1, \dots, n.
\label{eq:def_indep_iid}
\end{equation}
We assigned a Gamma hyperprior with parameters $1$ (shape) and $0.00005$ (inverse scale) to the inverse of $\sigma^2_{\theta_{1}}$ and $\sigma^2_{\theta_{2}}$, i.e. the precisions. 

The spatially structured term in the first baseline model was defined using an \textit{Independent Intrinsic Multivariate Conditional Auto-regressive} (IIMCAR) prior, whereas, for the second model, we adopted an \textit{Independent Proper Multivariate Conditional Auto-regressive} (IPMCAR) prior. 
The IIMCAR and IPMCAR distributions are briefly introduced hereafter, starting from their classical univariate counterparts, namely the ICAR and PCAR distributions. 

Univariate spatial random effects are traditionally modelled using a prior that belongs to the family of \textit{Conditional Auto-regressive} (CAR) distributions \parencite{besag1974spatial}. 
Given a random vector $\bm{\phi} = (\phi_1, \dots, \phi_n)$, the \textit{Intrinsic Conditional Auto-Regressive} (ICAR) distribution, which is a particular case of the CAR family, is usually defined through a set of conditional distributions \parencite{besag1995conditional}:
\begin{equation}
\phi_i | \lbrace \phi_{i'}, i' \in \partial_i \rbrace; \sigma^2 \sim N\left(m_i^{-1}\sum_{i' \in \partial_i}\phi_{i'}, \frac{\sigma^2}{m_i}\right), \ i = 1, \dots, n,
\label{eq:def_conditional_ICAR}
\end{equation}
where $\partial_i$ and $m_i$ denote, respectively, the indices and the cardinality of the set of neighbours for spatial unit $i$. 
These quantities are defined through a sparse binary symmetric neighbourhood matrix $\bm{W}$ with dimensions $n \times n$ that summarises the spatial relationships in the region of study. 
We built it taking advantage of the dual representation of a road network as a spatial and a graph object (see \textcite{porta2006network} and Section~\ref{sec:data}). 
More precisely, $\bm{W}$ is the adjacency matrix of a graph whose vertices correspond to the street segments of the road network and the edges identify a shared point at the boundaries of two spatial units.
This procedure defines a \textit{First Order} neighbourhood matrix. 
\textit{Second} and \textit{Third Order} neighbourhood matrices are defined iteratively in the same way.

It is possible to prove that the prior defined by~\eqref{eq:def_conditional_ICAR} suffers from rank-deficiency problems, that are usually fixed by imposing a set of sum-to-zero constraints on the vector $\bm{\phi}$, one for each group of connected segments in the graph of the road network \parencite{hodges2003precision}.  
In this paper, we deal with a fully connected road network (see the pre-processing procedures detailed in Section~\ref{sec:data}), so we always had to fix only one set of constraints.

The \textit{Proper Conditional Auto-Regressive} (PCAR) distribution is another member of the CAR family and it is usually defined as follows: 
\begin{equation}
\phi_i | \lbrace \phi_{i'}, i' \in \partial_i \rbrace; \sigma^2, \rho \sim N\left(\rho\left(m_i^{-1}\sum_{i' \in \partial_i}\phi_{i'}\right), \frac{\sigma^2}{m_i}\right), \ i = 1, \dots, n,
\label{eq:def_conditional_PCAR}
\end{equation}
where $\partial_i$ and $m_i$ are defined as for the ICAR distribution and $\rho$ is a parameter controlling the strength of spatial dependence, usually called \textit{spatial autoregression coefficient} \parencite{cressie1993statistics}. 
It is possible to prove that the joint distribution defined by~\eqref{eq:def_conditional_PCAR} is proper if $\abs{\rho} < 1$, hence there is no need to set any sum-to-zero constraint in this case. 
The ICAR prior can be seen as a limit case of the PCAR distribution with $\rho \to 1$, analogously to the relationship between Auto-Regressive and Random-Walk models in time series models \parencite{botella2013spatial}.

The family of \textit{Multivariate Conditional Auto-regressive} (MCAR) distributions was firstly introduced by \textcite{mardia1988multi}, extending the ideas of \textcite{besag1974spatial} to the  multivariate case. 
Given a random matrix $\bm{\Phi} = (\phi_{ij})$, which is defined for $i = 1, \dots, n$ units and $j = 1, \dots, J$ levels, the \textit{Intrinsic Multivariate Conditional Auto-regressive} (IMCAR) distribution is a particular case of the MCAR family, defined through a set of multivariate conditional distributions \parencite{martinez2019disease}: 
\begin{equation}
\bm{\Phi}_{i\cdot} | \text{vec}\left(\bm{\Phi}_{-i\cdot}\right); \Omega \sim N_{J}\left(m_i^{-1}\sum_{i' \in \partial_i}\bm{\Phi}_{i'\cdot}^{T}; m_i^{-1}\Omega^{-1}\right).
\label{eq:def_conditional_IMCAR}
\end{equation}
The terms $\bm{\Phi}_{i\cdot}$ and $\bm{\Phi}_{-i\cdot}$ denote, respectively, the $i$th row of $\bm{\Phi}$ and the matrix obtained by excluding the $i$th row from $\bm{\Phi}$. 
The $\text{vec}$ operator is used for row-binding the columns of a matrix, meaning that $\text{vec}\left(\bm{\Phi}_{-i\cdot}\right) = \left(\bm{\Phi}_{-i1}^{T}, \dots, \bm{\Phi}_{-iJ}^{T}\right)^{T}$. 
The elements $m_i$ and $\partial_i$ are defined as before, through the adjacency matrix $\bm{W}$ of the graph associated to the road network, and they represent the spatial dimension of the IMCAR distribution. 
The $J \times J$ precision matrix $\Omega$ is used to model the associations between pairs of levels in the same road segment $i$, and it acts as a multivariate extension of the parameter $\sigma^2$ in equation~\eqref{eq:def_conditional_ICAR}.

This distribution suffers from the same rank-deficiency problems as its univariate counterpart, which are usually solved by imposing appropriate sum-to-zero constraints. 
The number of restrictions is equal to the number of clusters in the graph of the road network times the number of levels in the multivariate setting. 
The pre-processing operations that we performed on the network data (see Section~\ref{sec:data}) imply that we always have to set only $J$ sum-to-zero constraints. 

The IIMCAR distribution is a particular case of~\eqref{eq:def_conditional_IMCAR}, which is obtained by setting $\Omega^{-1} = \text{diag}(\sigma^2_{\phi_1}, \dots, \sigma^2_{\phi_J})$. 
More precisely, if we assume $J = 2$ as we do in this paper, then IIMCAR is defined by the following set of multivariate conditional distributions: 
\begin{equation}
\bm{\Phi}_{i\cdot} | \text{vec}\left(\bm{\Phi}_{-i\cdot}\right); \sigma_{\phi_1}^2, \sigma_{\phi_2}^2 \sim N_{2}\left(m_i^{-1}\sum_{i' \in \partial_i}\bm{\Phi}_{i'\cdot}^{T}; m_i^{-1}
\begin{bmatrix}
\sigma_{\phi_1}^2 & 0 \\ 
0 & \sigma_{\phi_2}^2
\end{bmatrix}\right).
\label{eq:def_conditional_IIMCAR}
\end{equation}
In equation~\eqref{eq:def_conditional_IIMCAR} we are assuming independence between the $2$ levels, and this implies that the IIMCAR distribution is equivalent to two independent ICAR distributions, one for each level.

Analogously to the univariate case, the \textit{Proper Multivariate Conditional Auto-regressive} (PMCAR) distribution is a particular case of the MCAR family characterised by the following set of multivariate conditional distributions: 
\begin{equation}
\bm{\Phi}_{i\cdot} | \text{vec}\left(\bm{\Phi}_{-i\cdot}\right); \rho, \Omega \sim N_{J}\left(m_i^{-1}\rho\sum_{i' \in \partial_i}\bm{\Phi}_{i'\cdot}^{T}; m_i^{-1}\Omega^{-1}\right) 
\label{eq:def_conditional_PMCAR}. 
\end{equation}
The strength of the spatial dependence is controlled by $\rho$ (as for the univariate PCAR distribution) and all the other parameters are defined as before. 
It can be proved that the joint distribution defined by equation~\eqref{eq:def_conditional_PMCAR} is proper if $\abs{\rho} < 1$, although we restricted ourself to $\rho \in (0, 1)$ to avoid some counter-intuitive behaviour of the PMCAR distribution \parencite{wall2004close, miaou2005bayesian}. 

The IPMCAR distribution is defined as a particular case of equation~\eqref{eq:def_conditional_PMCAR} with $\Omega^{-1} = \text{diag}(\sigma^2_{\phi_1}, \dots, \sigma^2_{\phi_J})$. 
More precisely, if we assume $J = 2$, then IPMCAR is defined through the following set of multivariate conditional distribution: 
\begin{equation}
\bm{\Phi}_{i\cdot} | \text{vec}\left(\bm{\Phi}_{-i\cdot}\right); \rho, \sigma^2_{\phi_1}, \sigma^2_{\phi_2} \sim N_{2}\left(m_i^{-1}\rho\sum_{i' \in \partial_i}\bm{\Phi}_{i'\cdot}^{T}; m_i^{-1}\begin{bmatrix}
\sigma^2_{\phi_1} & 0 \\ 
0 & \sigma^2_{\phi_2}
\end{bmatrix}\right).
\label{eq:def_conditional_IPMCAR}
\end{equation}
For the same reasoning as in equation~\eqref{eq:def_conditional_IIMCAR}, the IPMCAR distribution is equivalent to $J$ independent PCAR distributions.

Now we can characterise the random effects for the two baseline models. 
The first model was defined by considering unstructured random effects with a bivariate independent Gaussian prior \eqref{eq:def_indep_iid}, and spatial random effects with an IIMCAR prior \eqref{eq:def_conditional_IIMCAR}. 
The second one was defined analogously to the first baseline model, but assuming an IPMCAR distribution for the spatial random effects \eqref{eq:def_conditional_IPMCAR}. 
These models assume independence between the two levels both in the spatial and unstructured components, so they were used as benchmarks. 
In the next sections, we will also refer to the two baseline models using, respectively, the codes (A) and (B). 
We assigned an improper prior to $\sigma^2_{1}$ and $\sigma^{2}_{2}$, the variances in $\Omega$, defined on $\mathbb{R}^{+}$, and a $\text{Uniform}(0, 1)$ prior to $\rho$.

Hereafter we introduce two increasingly complex sets of extensions that generalise the baseline models. 
The first one is characterised by the removal of the independence assumption from the spatially structured random effects, whereas, in the second set of extensions, we also relax the independence assumption from the unstructured random effects. 

\subsection{Model extensions}

\subsubsection*{First set of extensions}

Starting from the baselines, we defined two new models replacing the IIMCAR and IPMCAR priors with their non-independent multivariate counterparts, the generic IMCAR and PMCAR defined above.
If we assume $J = 2$, then the variance-covariance matrix $\Omega^{-1}$ in \eqref{eq:def_conditional_IMCAR} and \eqref{eq:def_conditional_PMCAR} can be written as 
\[
\Omega ^ {-1} = \begin{bmatrix}
\sigma^2_{\phi_1} & \rho_{\phi}\sigma_{\phi_1}\sigma_{\phi_2} \\
\rho_{\phi}\sigma_{\phi_1}\sigma_{\phi_2} & \sigma^2_{\phi_2}
\end{bmatrix},
\] 
where $\sigma_1^2$ and $\sigma_2^2$ represent the conditional variances and $\rho_{\phi}$ represents the correlation coefficient between the two levels in the same spatial unit.
These models represent a generalisation of the baselines since we are now taking into account the correlations between different levels in the same road segment. 
We will also refer to them using, respectively, the codes (C) and (D). 
Following \textcite{palmiperales2019bayesian}, we assigned a Wishart hyperprior to $\Omega^{-1}$ with parameters $2$ and $\bm{I}_{2}$,  i.e. the identity matrix of size two. 
The prior distributions on the unstructured random effects were left unchanged with respect to the baselines.

\subsubsection*{Second set of extensions}

In these models, the independence assumption of the spatially unstructured random effects is removed. 
More precisely, assuming $J = 2$, we assign a generic bivariate Gaussian prior to the unstructured random effects: 
\[
\left(\theta_{i1}, \theta_{i2}\right)\sim N_2\left(\bm{0}, \begin{bmatrix}
\sigma^{2}_{\theta_{1}} & \rho_{\theta}\sigma_{\theta_{1}}\sigma_{\theta_{2}} \\
\rho_{\theta}\sigma_{\theta_1}\sigma_{\theta_2} & \sigma^2_{\theta_{2}}
\end{bmatrix}\right). 
\]
Parameters $\sigma_{\theta_{1}}^2$ and $\sigma_{\theta_{2}}^2$ represent the marginal variances of the unstructured random error, whereas $\rho_{\theta}$ represents their correlation. 
These models will also be identified using the codes (E) and (F). 
We assigned a Wishart hyperprior to the variance-covariance matrix with parameters $2$ and $\bm{I}_2$. 

Finally, following the ideas introduced in \textcite{gelfand2003proper}, we also tested an extension of model (F), named model (G), which is characterised by a generalisation of the PMCAR distribution that introduces a separate spatial autoregression coefficient, $\rho_{j}$, for each level in $\bm{\Phi}$. 
We found that this extension did not improve over model (F). 
Hence, we will not add more details here and refer the interested reader to the supplementary materials.

The prior distributions adopted for the random effects in the two baselines and their extensions are summarised in Table~\ref{tab:summary_models}. 
We also included the IDs that will be used to identify each model in subsequent Tables and Sections. 

\begin{table}
	\centering
	\caption{Summary of the prior distributions assigned to the random effects in the models introduced in Section~\ref{sec:methodology}.}
	\begin{tabular}{llll}
		\toprule
		ID & Model & Unstructured Effect & Spatial effect\\
		\midrule 
		(A) & Baseline 1  & Independent Gaussian & Independent IMCAR \\
		(B) & Baseline 2  & Independent Gaussian & Independent PMCAR \\
		(C) & Extension 1 - Model 1 & Independent Gaussian & IMCAR \\
		(D) & Extension 1 - Model 2 & Independent Gaussian & PMCAR \\
		(E) & Extension 2 - Model 1 & Correlated Gaussian  & IMCAR \\
		(F) & Extension 2 - Model 2 & Correlated Gaussian  & PMCAR \\
		\bottomrule
	\end{tabular}
	\label{tab:summary_models}
\end{table}

\subsection{Model comparison}

The models proposed in the previous paragraphs were compared using Deviance Information Criterion (DIC) \parencite{spiegelhalter2002bayesian} and Watanabe–Akaike Information Criterion (WAIC) \parencite{gelman2014understanding, watanabe2010asymptotic}. 
These criteria represent a measure for the adequacy of a model, penalised by the number of effective parameters. 
In both cases, the lower is the value of the index, the better is the fitting of the model.

\section{Results}
\label{sec:results}

We estimated the models previously described using the software \texttt{INLA} \parencite{INLA1, INLA2, gomez2020bayesian}, interfaced through the homonymous R package \parencite{RSoftware}.
We used the \textit{Simplified Laplace} strategy for approximating the posterior marginals and the \textit{Central Composite Design} strategy for determining the integration points. 
The code behind the definition of multivariate ICAR and PCAR random effects is defined in the package \texttt{INLAMSM} \parencite{palmiperales2019bayesian}. 
It took approximately 30 - 45 minutes to estimate each model using a virtual machine with an Intel Xeon E5-2690 v3 processor, six cores, and 32GB of RAM. 

\subsubsection*{Fixed effects}

The models listed in Table~\ref{tab:summary_models} share a common structure for the fixed effect component, with a severity-specific intercept and a set of covariates representing some social and physical characteristics of the road segments. 
In particular, we considered five severity-specific covariates, namely the two socio-economic variables mentioned in Section~\ref{sec:data}, a dummy variable recording whether a road segment lays in a dual carriageway street or not, the road type (either \textit{Motorway}, \textit{Primary Road} or \textit{A Road}, according to OS definition), and the edge betweenness centrality measure, which reflects the number of shortest paths traversing each segment \parencite{kolaczyk2014statistical}. 
The last covariate can be considered as a proxy for the average vehicle miles traveled (VMT) \parencite{briz2019estimating}, the population density can be thought of as a surrogate for pedestrian and cycling volumes, and the employment rates may represent an indirect measure of income.  
Finally, to improve the stability of INLA algorithms, we scaled all numerical variables to zero mean and unit variance.

Table~\ref{tab:fixed_effects} shows the posterior means and standard deviations for the fixed effects. 
We first notice that the estimates are stable among the models. 
The intercept for severe car crashes, $\beta_{01}$, is found slightly smaller than $\beta_{02}$. 
This is not surprising since severe accidents are rarer than the slight ones. 
The coefficients of edge betweenness centrality measures are found close to zero for all models, and their 95\% credible interval (not reported in the table) always include the value zero.  
The Road type parameters represent relative differences with respect to the reference category (i.e. \textit{A Roads}), hence \textit{Motorways} are found less prone to severe and slight car crashes than \textit{A roads}. 
A similar finding was also reported by \textcite{boulieri2017space} for UK data.
An analogous interpretation applies to \textit{Primary Roads}. 
In agreement with other studies (see, for instance \parencite{HUANG201710} and references therein), we found that population density significantly correlates to both slight and severe accidents, although its effect is stronger for the former type.
On the other hand, the second socio-economic variable was not found significant in any of the outcomes considered in this paper. 
Unfortunately, more direct income measures or poverty indicators are not available and are difficult to construct at such detailed spatial resolution. 
More work is needed to to assess the relationship between road safety and socio-economic inequalities.
Finally, dual carriageway roads have been found significantly less prone to slight car accidents, whereas no impact has been found for severe car crashes. 
An analogous result was also reported by \textcite{star-rating-UK}.  

\setlength{\tabcolsep}{1pt}

\begin{sidewaystable}
	\centering
	\caption{Estimates for the posterior means and standard deviations, in round brackets, of the fixed effects included in the models described in Table~\ref{tab:summary_models}.}
	{
		\small
		\begin{tabular}{l|SSSSSSS|SSSSSSS} 
			\toprule
			\multirow{2}{*}{ID} & \multicolumn{7}{c}{Severe Crashes} & \multicolumn{7}{c}{Slight Crashes}\\
			& {$\beta_{01}$} & {Betw.} & {Motorways} & {Prim. Roads} & {Ratio Empl.} & {Pop. Dens.} & {Dual Carr.} & {$\beta_{02}$} & {Betw.} & {Motorways} & {Prim. Roads}  & {Ratio Empl.} & {Pop. Dens.} & {Dual Carr.} \\
			\midrule 
			\multirow{2}{*}{(A)} 
			& -14.229 &  0.011 & -0.850 & 0.327 & 0.009  & 0.181 & 0.040 & 
			-12.679 & -0.004 & -0.009 & 0.656 & -0.005 & 0.265 & -0.318 \\ 
			& (0.092) & (0.058) & (0.180) & (0.148) & (0.061) & (0.053) & (0.110) & 
			(0.061) & (0.036) & (0.117) & (0.101) & (0.043) & (0.040) & (0.072) \\
			\multirow{2}{*}{(B)} 
			& -14.348 & -0.015 & -0.834 & 0.336 & 0.026 & 0.157 & 0.031 &
			-12.700 & -0.023 & -0.023 & 0.630 & 0.011 & 0.258 & -0.324 \\ 
			& (0.153) & (0.059) & (0.187) & (0.150) & (0.063) & (0.055) & (0.114) & 
			(0.132) & (0.036) & (0.123) & (0.104) & (0.045) & (0.042) & 0.073 \\
			\multirow{2}{*}{(C)} 
			& -14.245 & -0.051 & -0.737 & 0.397 & 0.011 & 0.168 & -0.023
			& -12.665 & -0.031 & 0.011 & 0.636&-0.005&0.241&-0.317
			\\ 
			&(0.092)&(0.058)&(0.185)&(0.149)&(0.063)&(0.056)&(0.108)
			&(0.062)&(0.036)&(0.121)&(0.104)&(0.045)&(0.041)&(0.071)
			\\
			\multirow{2}{*}{(D)} 
			&-14.354&-0.069&-0.697&0.516&0.008&0.115&-0.111 
			&-12.662&-0.038&-0.077&0.58&0.022&0.243&-0.3
			\\ 
			&(0.12)&(0.053)&(0.181)&(0.131)&(0.059)&(0.055)&(0.107)
			&(0.098)&(0.036)&(0.124)&(0.101)&(0.045)&(0.042)&(0.072)
			\\
			\multirow{2}{*}{(E)}
			&-14.285&-0.005&-0.805&0.426&-0.007&0.209&-0.006 
			&-12.665&-0.012&-0.021&0.639&-0.004&0.261&-0.299
			\\ 
			&(0.092)&(0.057)&(0.183)&(0.143)&(0.062)&(0.056)&(0.11)
			&(0.061)&(0.036)&(0.117)&(0.101)&(0.043)&(0.04)&(0.071)
			\\
			\multirow{2}{*}{(F)} 
			& -14.357&-0.043&-0.758&0.473&0.009&0.155&-0.053
			& -12.677&-0.026&-0.055&0.599&0.015&0.256&-0.299
			\\ 
			&(0.138)&(0.055)&(0.183)&(0.136)&(0.06)&(0.056)&(0.11)
			&(0.119)&(0.036)&(0.121)&(0.102)&(0.044)&(0.041)&(0.072)
			\\
			\bottomrule
		\end{tabular}
	}
	\setlength{\tabcolsep}{6pt}
	\vspace{1.5cm}
	\label{tab:fixed_effects}
	\caption{Estimates for the posterior means and standard deviations, in round brackets, of hyperparameters included in the models described in Table~\ref{tab:summary_models}.}
	\begin{tabular}{l|SSSSSSS}
		\toprule {ID} & {$\sigma^2_{\theta_1}$} & {$\sigma^2_{\theta_2}$} & {$\rho_\theta$} & {$\rho$} & {$\sigma^2_{\phi_1}$} & {$\sigma^2_{\phi_2}$} & {$\rho_\phi$} \\
		\midrule 
		\multirow{2}{*}{(A)} & 0.0001 & 0.759 & & & 0.152 & 0.127 & \\
		& (0.0003) & (0.049) & & & (0.028) & (0.019) & \\
		\multirow{2}{*}{(B)} & 0.0001 & 0.578 & & 0.996 & 0.373 & 0.347 & \\
		& (0.0003) & (0.047) & & (0.0008) & (0.052) & (0.045) &  \\
		\multirow{2}{*}{(C)} & 0.0001 & 0.594 & & & 0.308 & 0.249 & 0.898 \\
		& (0.0003) & (0.048) & & & (0.047) & (0.035) & (0.021) \\
		\multirow{2}{*}{(D)} & 0.0001 & 0.376 & & 0.986 & 0.747 & 0.690 & 0.904 \\
		& (0.0003) & (0.054) & & (0.005) & (0.121) & (0.108) & (0.019) \\
		\multirow{2}{*}{(E)} & 0.589 & 0.752 & 0.413 & & 0.132 & 0.134 & 0.789 \\
		& (0.081) & (0.046) & (0.013) & & (0.020) & (0.017) & (0.042) \\		
		\multirow{2}{*}{(F)} & 0.484 & 0.643 & 0.405 & 0.997 & 0.276 & 0.269 & 0.829 \\
		& (0.081) & (0.046) & (0.016) & (0.001) & (0.038) & (0.033) & (0.030) \\ 
		\bottomrule
	\end{tabular}
	\label{tab:random_effects}
\end{sidewaystable} 

\subsubsection*{Random effects}

The posterior means and standard deviations for all hyperparameters are reported in Table~\ref{tab:random_effects}, which reflects the models nested structure, also summarised in Table~\ref{tab:summary_models}. 
We started from two baselines, (A) and (B), with independent random effects, and generalised them until model (F), that presents multiple autocorrelation parameters between the two severity levels.

Models from (A) to (D), which assume independent unstructured random effects, exhibit a degenerate posterior distribution of $\sigma_{\theta_{1}}^2$, i.e. the variance of severe random component, and this is possibly due to the severe car crashes sparseness. 
This problem gets mitigated once the correlation parameter between the two severity levels is included in the model, suggesting that the estimation procedure benefits from the inclusion of a multivariate structure that allows borrowing strength from less rare events. 

The estimates of $\sigma_{\theta_{2}}^2$ and $\rho_{\theta}$ are stable among the models, and the correlation parameter is estimated as high as $0.40$, suggesting a positive and mildly strong relationship between the two random components.  
The posterior means for hyperparameters $\rho$  in models (B), (D), and (F), are always very close to one, which is not uncommon for this type of models \parencite{carlin2003hierarchical}. 
The estimates of the posterior distributions for the two conditional variances, $\sigma_{\phi_1}^2$ and $\sigma_{\phi_2}^2$, are found less stable compared to the unstructured errors. 
The credible intervals of the two hyperparameters overlap in all models, indicating a similar spatial structure between the two kinds of severities. 
The posterior mean of $\rho_\phi$, the correlation coefficient between the two severity levels, is found approximately equal to 0.9 (models C and D) and 0.8 (models E and F), indicating a strong multivariate nature for the spatial random component.

These results suggest that car crash data have a complex latent structure being the severity levels strongly correlated, and the spatially structured and unstructured effects statistically relevant. 

\subsubsection*{Model comparisons}

We compared the models listed in Table~\ref{tab:summary_models} using DIC and WAIC criteria. 
The results are reported in Table~\ref{tab:DIC}. 
We first notice that PMCAR models (i.e. (B), (D) and (F)) are found to perform always better than their Intrinsic counterparts in terms of goodness of fit. 
They are somewhat unexplored in the road safety literature on spatial networks, \textcite{miaou2005bayesian} being the only paper we found that analyse the importance of a spatial autocorrelation parameter. 
However, our results suggest that PCAR distribution and its generalisations should deserve more attention. 

Moving from (A) to (F) the model performance improves, indicating one more time the benefits of considering a correlated multivariate structure for the spatial and the unstructured components. 
In particular, model (F) is the best one according to both criteria; hence, hereafter, we focus on this model.  

\begin{table}
	\centering
	\caption{Estimates of DIC and WAIC values for the models described in Section~\ref{sec:methodology}. The columns \textit{Balanced Accuracy-Severe} and \textit{Balanced Accuracy - Slight} will be explained in Section~\ref{sec:criticism}.}
	\begin{tabular}{lrrrr}
		\toprule ID & DIC & WAIC & Balanced Accuracy - Severe & Balanced Accuracy - Slight \\
		\midrule 
		(A) & 14462.56 & 14474.89 & 0.631 & 0.718 \\
		(B) & 14408.06 & 14433.77 & 0.635 & 0.716 \\
		(C) & 14269.79 & 14296.06 & 0.645 & 0.718 \\
		(D) & 14154.09 & 14167.64 & 0.661 & 0.717 \\
		(E) & 14122.46 & 14106.69 & 0.662 & 0.720 \\
		(F) & 14103.44 & 14086.46 & 0.675 & 0.720 \\
		\bottomrule
	\end{tabular}
	\label{tab:DIC}
\end{table}

\subsubsection*{Car crashes rates}

\begin{figure}
	\centering
	\subfloat[Severe Car Crashes. \label{fig:pred_serious}]{
		\includegraphics[width=\linewidth]{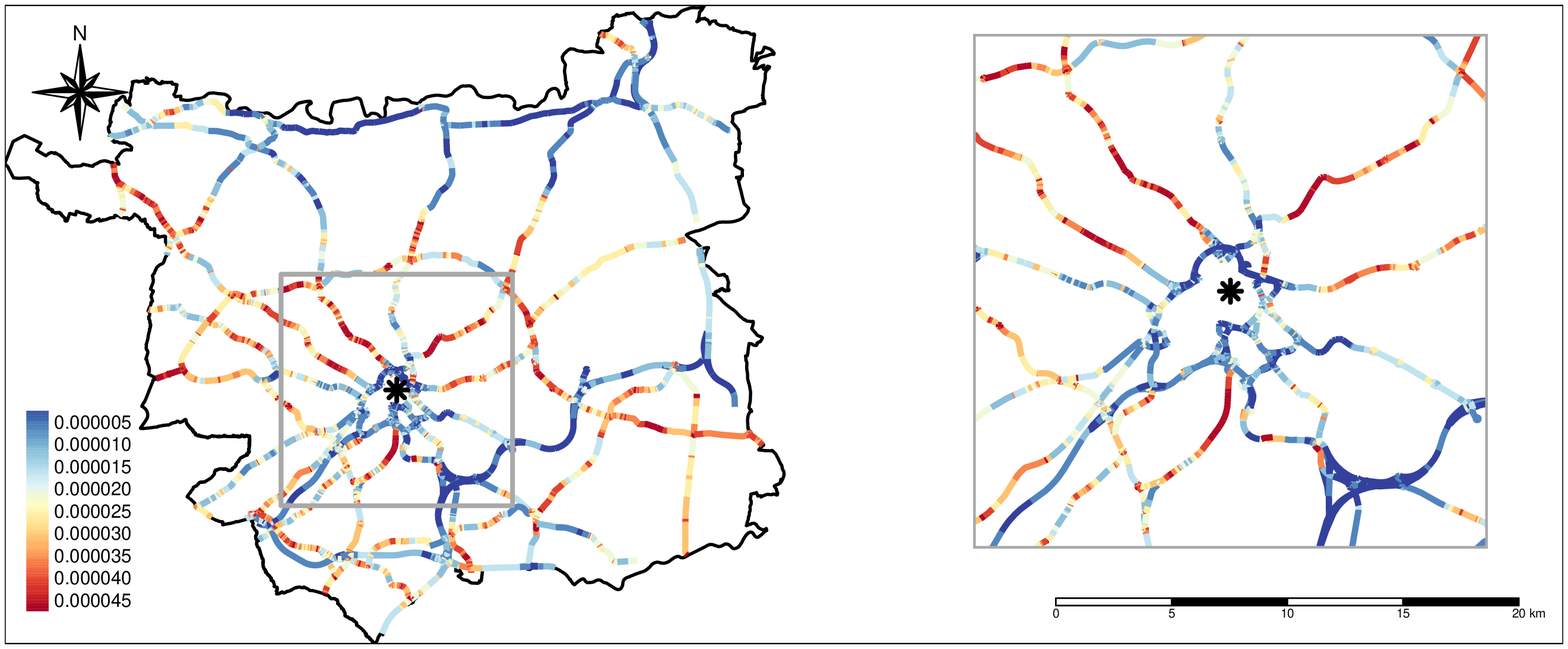}
	}
	\hspace{0.01\linewidth}
	\subfloat[Slight Car Crashes. \label{fig:pred_slight}]{
		\includegraphics[width=\linewidth]{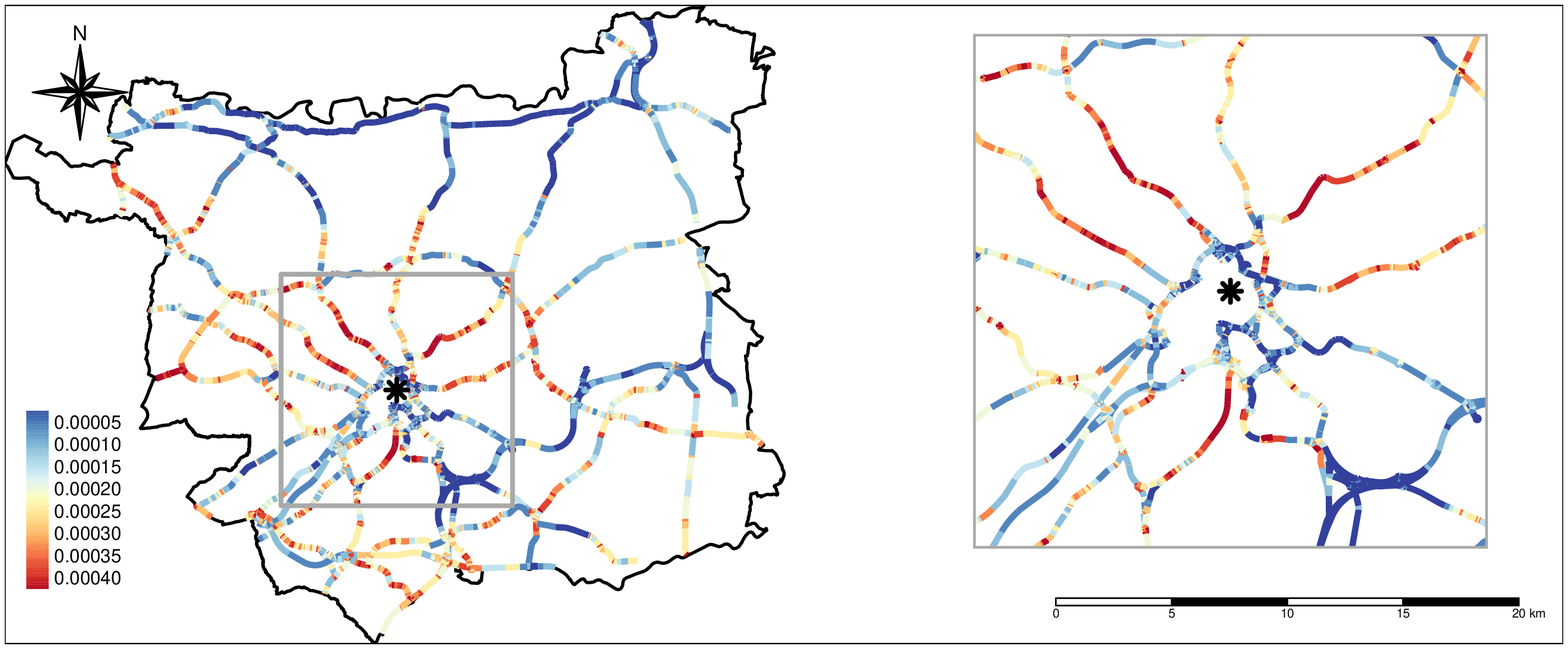}
	}
	\caption{Maps representing the posterior means for severe and slight car crashes rates, estimated using model (F). The colours go from red (higher quantiles) to blue (lower quantiles). The black star indicates the city centre while the inset maps highlight the road network in its proximity.}
	\label{fig:pred}
\end{figure}

Figure~\ref{fig:pred} displays the posterior means of car accident rates, $\lambda_{ij}$, estimated using model (F) both for severe and slight crashes.  
The colours of the road segments were generated by dividing the predicted values of each severity level into ten classes based on a set of quantiles ranging from red (highest quantile) to blue (lowest quantiles). 
In both cases, the highest values correspond to, approximately, one severe or ten slight car crashes every kilometre and every thousand of daily commuters (i.e. the two quantities that define the offset). 
Moreover, the inset maps highlight a few roads close to Leeds city centre (denoted by a black star). 

The four maps show similar patterns, but some roads in the southern part of the city (especially M621) look more prone to severe car crashes. 
The city of Leeds appeared to be divided into several areas. 
The northern and north-eastern part of the city are associated with lower car accident rates compared to other suburbs. 
The areas located in north-west, east, and south of the city centre seem to be associated with the highest levels of car crashes rates, especially severe ones. 
This is probably linked with some congested arterial thoroughfares reaching the centre.
Finally, we note that the roads closer to the city centre are the safest part of the city network, as it is clearly shown by the two inset maps in Figure~\ref{fig:pred}.

\section{Model criticism and sensitivity analysis}
\label{sec:criticism}

DIC and WAIC criteria were never intended to be absolute measures of model fit, and they cannot be used for \textit{Model Criticism}. 
Hence, we tested the adequacy of model (F) using two strategies.  

\subsection{First strategy for criticism}

The classical criterion for criticism of a Bayesian hierarchical model is the Probability Integral Transform \parencite{marshall2003approximate, held2010posterior}, typically adjusted in case of a discrete response variable (such as car crashes counts) using a continuity correction. 
Unfortunately, these adjustments do not seem to work appropriately when modelling sparse count data, such as severe crashes, since the correction is not adequate.  
We refer the interested reader to the supplementary material for more details.

Therefore, hereafter, we followed a different strategy. 
We binned the observed and predicted counts into two classes: \textit{Zero} and \textit{One or more} car crashes.
Then, we built a confusion matrix and evaluated the model performance via some accuracy measures that are summarised in Table~\ref{tab:confusion_matrix}. 
A similar procedure for sparse count data was also presented in \textcite{ma2017multivariate}. 
We decided to adopt one as a threshold to dichotomise the variables since more than 80\% of road segments registered no severe car crash during 2011-2018. 
However, the algorithm proposed here can be extended to three or more classes, defined using a set of different thresholds (such as \textit{Zero}, \textit{One}, and \textit{Two or more} road crashes).

The \textit{accuracy} measure, usually adopted for evaluating the predictive performance of a model, is typically biased and overly-optimistic in case of unbalanced classes (such as \textit{Zero} and \textit{One or more} severe car crashes per road segment), since, even in the worst case, it is as high as the percentage of observations in the more frequent class \parencite{5597285}. 
The \textit{balanced accuracy}, firstly introduced by \textcite{5597285}, is defined as the average of Sensitivity and Specificity, and it overcomes this drawback since it represents an average between the predictive performances on each class. 

\begin{table}
	\centering
	\hspace{-2cm}
	\caption{Left: Confusion matrix showing the observed and predicted counts, binned in two classes. The rows represent the actual counts while the columns are the predicted counts. Right: Definition of accuracy measures.}
	\begin{tabular}{cccc}
		\multicolumn{2}{c}{} & Zero & One or more \\
		\multicolumn{2}{l}{Zero} & A & B \\
		\multicolumn{2}{l}{One or More} & C & D \\
	\end{tabular}
	\qquad
	\qquad 
	\begin{tabular}{ll}
		Sensitivity & = $\frac{A}{A + B}$ \\[3pt]
		Specificity & = $\frac{D}{C + D}$ \\[3pt]
		Accuracy & = $\frac{A + D}{A + B + C + D}$ \\[3pt]
		Balanced Accuracy & = $\frac{1}{2}\left(\frac{A}{A + B} + \frac{D}{C + D}\right)$
	\end{tabular}
	\label{tab:confusion_matrix}
\end{table}

The output of a Bayesian hierarchical model is an estimate of the posterior distribution of predicted values, while the procedure reported in the previous paragraph can only be applied to binary data. 
For this reason, we simulated $n$ Poisson random variables (one for each road segment) with mean equal to the mean of each posterior distribution. 
Then, we binned the observed and sampled counts into two classes, i.e. Zero and One or more car crashes, and we compared the two values, obtaining a single estimate of balanced accuracy. 
Its distribution was finally approximated by repeating this procedure $N = 5000$ times.

\begin{figure}
	\centering
	\subfloat[Severe Crashes \label{fig:balancedaccuracyserious}]{
		\includegraphics[width=0.47\linewidth, trim={0 0 0 0.735cm}, clip]{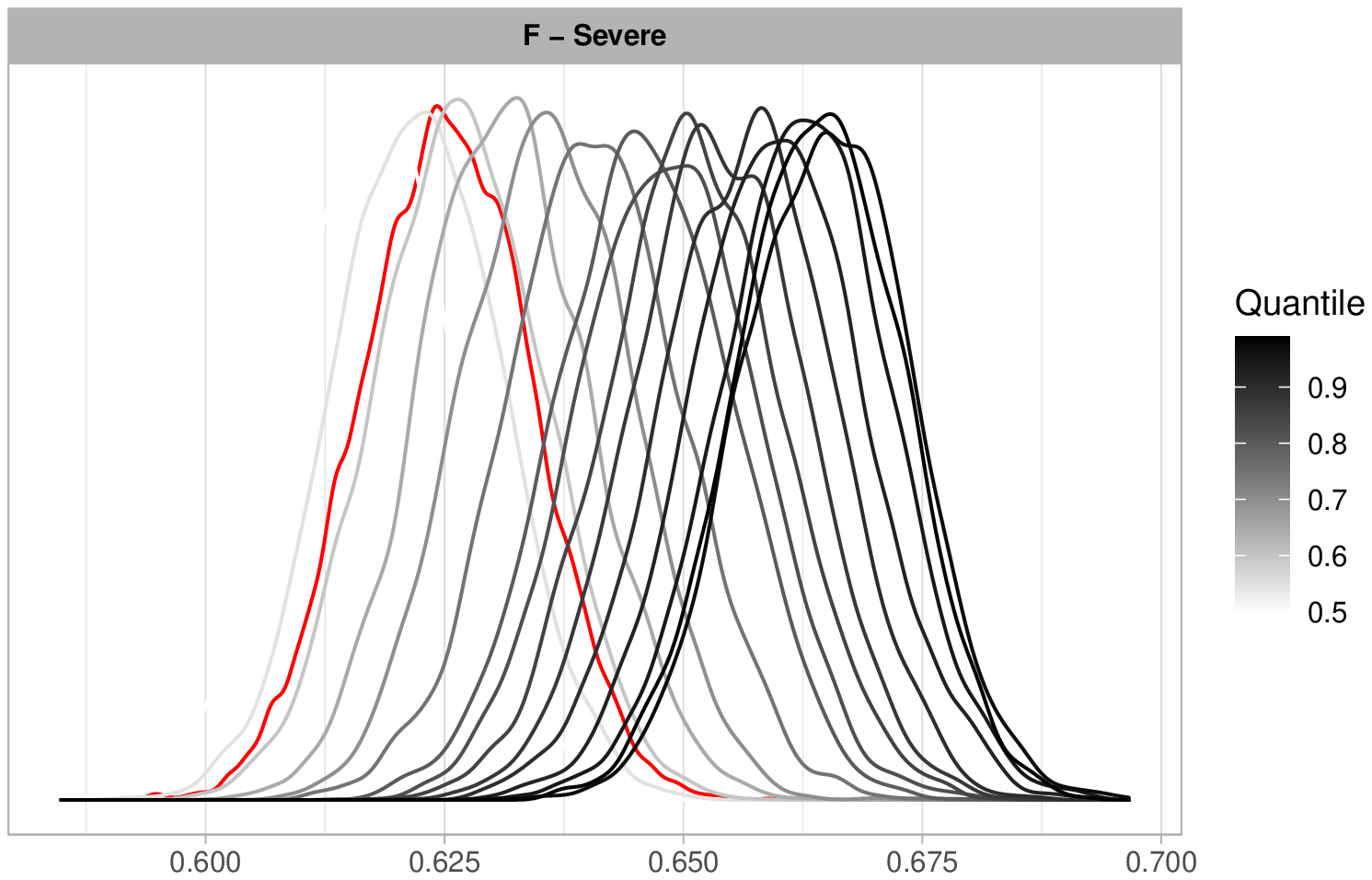}
	}
	\hspace{0.01\linewidth}
	\subfloat[Slight Crashes \label{fig:balancedaccuracyslight}]{
		\includegraphics[width=0.47\linewidth, trim={0 0 0 0.735cm}, clip]{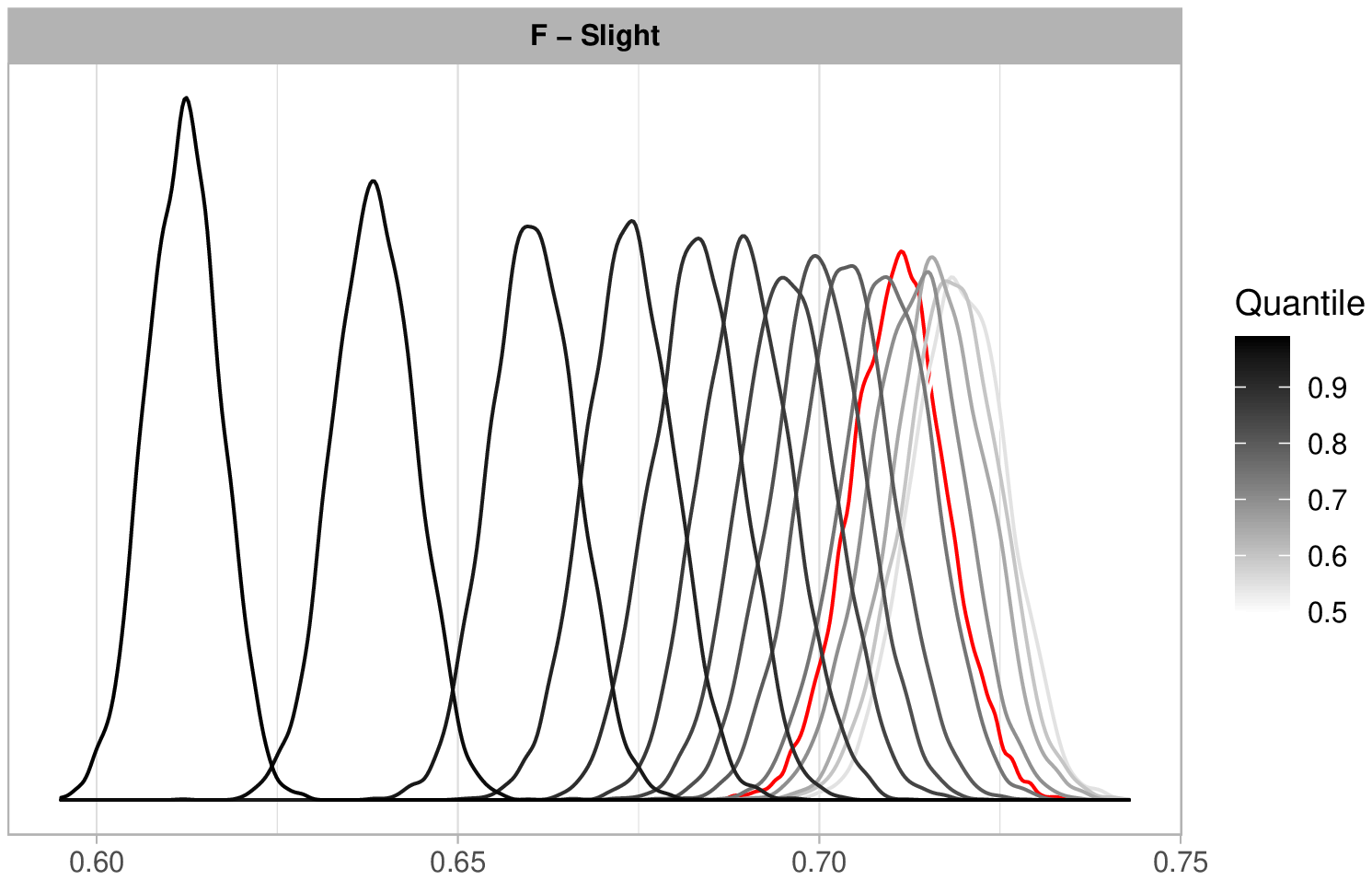}
	}
	\caption{Distribution of balanced accuracy for severe crashes (left) and slight crashes (right), considering a binary classification using the posterior mean and a set of quantiles. The red curve represents the mean.}
	\label{fig:balancedaccuracy}
\end{figure}

Moreover, we calculated several quantiles of the posterior distribution of each predicted value, and we run the same steps as before, sampling from a Poisson distribution with mean equal to each of those quantiles.  
Lastly, being severe and slight car crashes potentially quite different processes, this algorithm was applied independently for the two severity levels. 
We reported in the supplementary material the pseudo-code for running this procedure, whereas results are displayed in Figure~\ref{fig:balancedaccuracyserious} (severe cashes) and Figure~\ref{fig:balancedaccuracyslight} (slight crashes).  
In both cases, the red curve represents the distribution of balance accuracy obtained by a binary classification based on the posterior means, whereas the other curves represent the same distribution obtained using the set of quantiles. 

It looks like the optimal threshold for binary classification of severe car crashes is given by the $0.975$-quantile, where the balanced accuracy distribution is concentrated around $0.675$. 
The optimal threshold for binary classification of slight car crashes is given by the median, and the distribution of balanced accuracy is centred around $0.72$. 
These plots remark the differences between the two severity levels in terms of sparsity, suggesting the adoption of a higher quantile for the prediction of the more sparse events. 
However, using the appropriate cut-off(s), Model (F) seems to perform reasonably well in both cases with slightly better performance for slight car crashes. 
We also explored the goodness of fit of model (F) estimating the balanced accuracy measure separately for the three road classes and the two road carriageway types. 
We always obtained results similar to Figures~\ref{fig:balancedaccuracyserious} and~\ref{fig:balancedaccuracyslight}. 
Moreover, we did not observe any relevant difference among the road typologies, highlighting that model (F) successfully predicts traffic collisions in all tested situations.  

The third column in Table~\ref{tab:DIC} summarises the estimates of Balanced Accuracy obtained for each model in Table~\ref{tab:summary_models} under the best scenario (i.e. the optimal quantile) for severe crashes, while the fourth column summarises the same quantities for slight crashes. 
We notice that the accuracy improves every time a new correlation term is included into a model, particularly for severe car accidents, which is the rarer car crash typology. 
In fact, considering that improving risk estimation of very rare events is one of the main reasons why one may want to adopt a multivariate model in the first place, our results seem to suggest that the approach proposed in this paper represents a reasonable way to investigate road collision dynamics.

\subsection{Second strategy for criticism}

\begin{figure}
	\centering
	\subfloat[Severe Crashes \label{fig:posterior_means_serious}]{
		\includegraphics[width=0.47\linewidth]{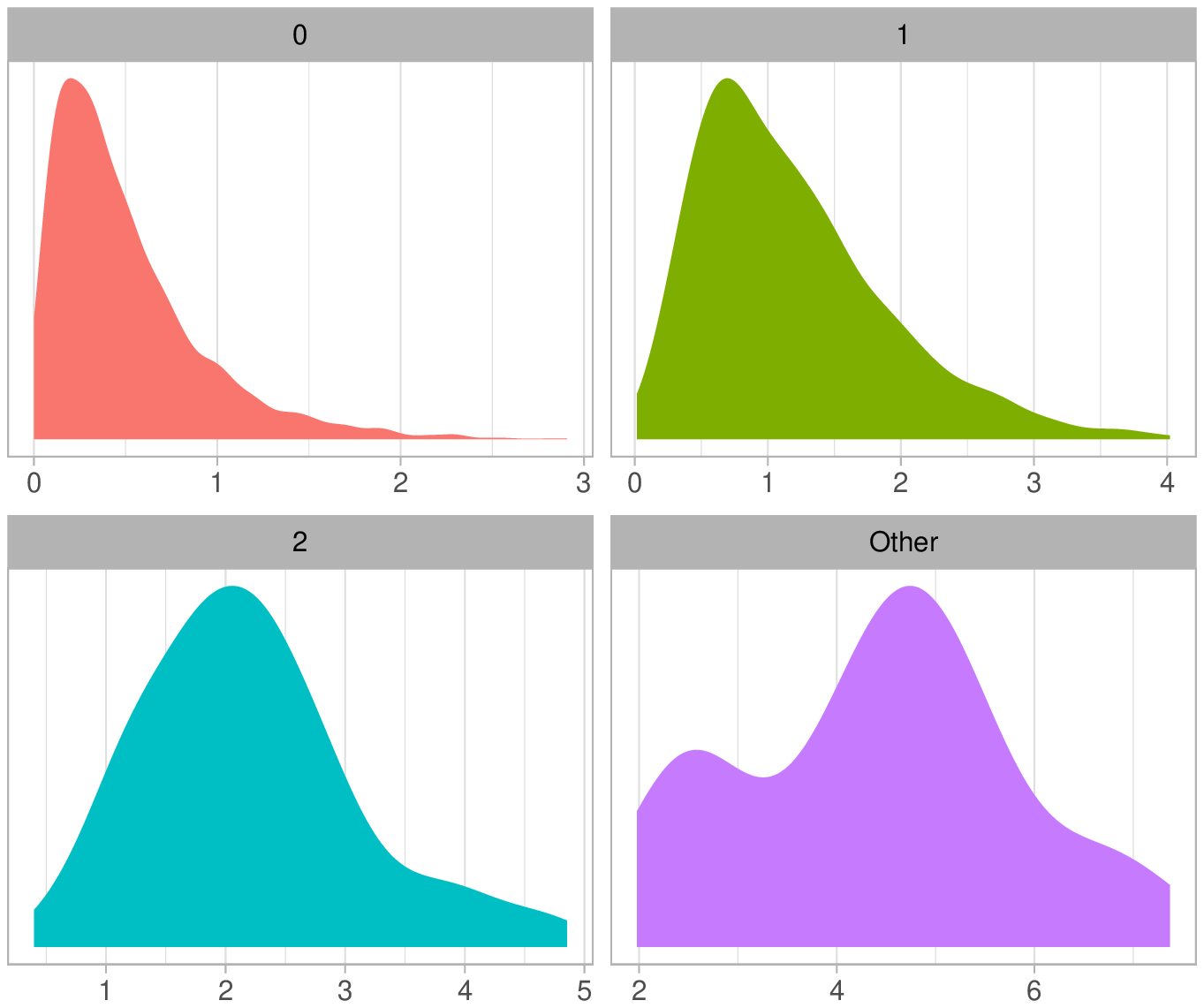}
	}
	\hspace{0.01\linewidth}
	\subfloat[Slight Crashes \label{fig:posterior_means_slight}]{
		\includegraphics[width=0.47\linewidth]{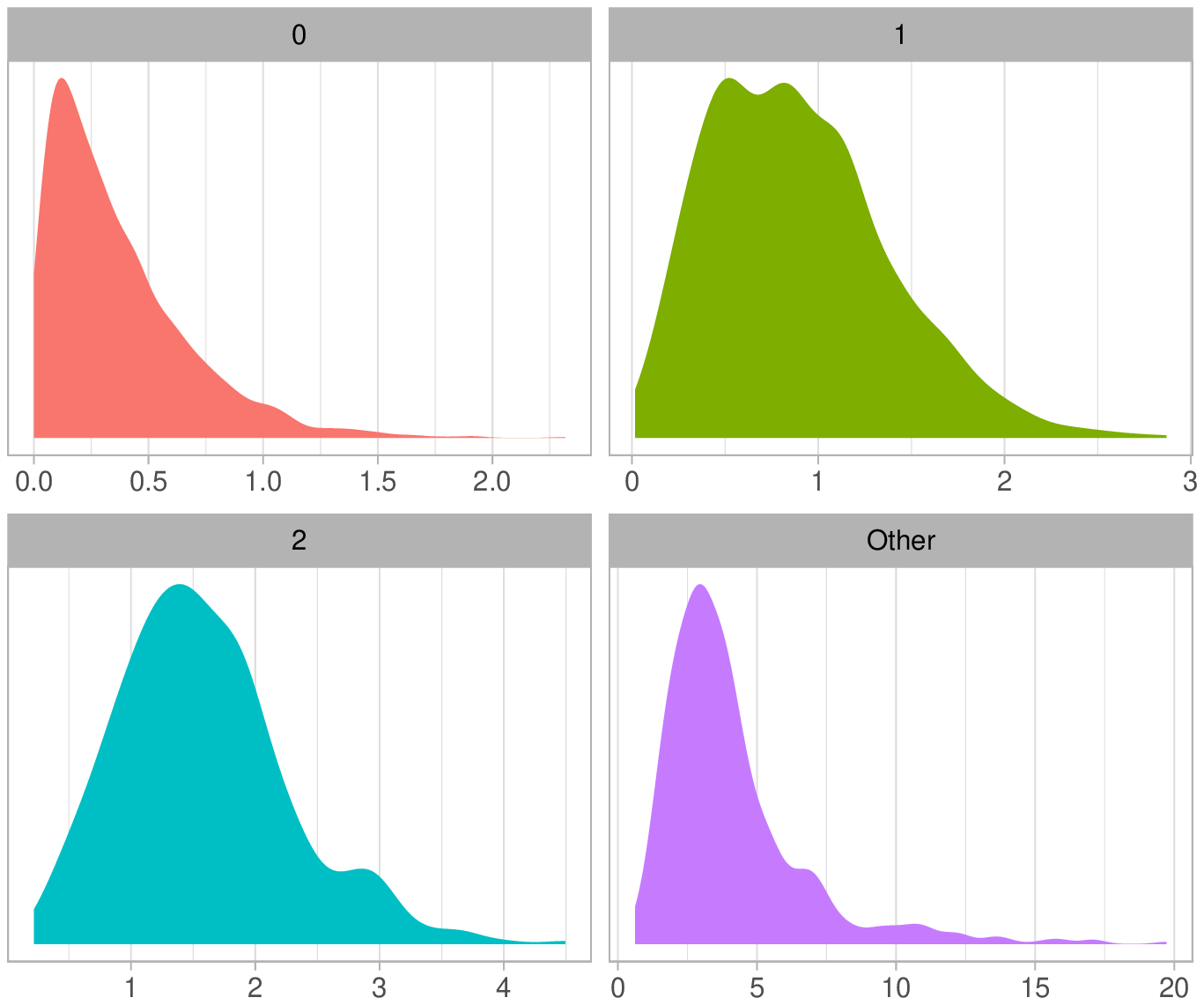}
	}
	\caption{Histogram of posterior $0.975$-quantile (left) and posterior median (right), grouped by the corresponding observed counts. Other means "Three or more."}
	\label{fig:posterior_means}
\end{figure}

Following the results illustrated before, we estimated the $0.975$-quantile of $\lambda_{i1}$ (severe crashes) and the median of $\lambda_{i2}$ (slight crashes), and we multiplied them by the corresponding offset values, i.e. $E_i$. 
Then, we created a sequence of histograms of predicted values, grouped by the observed counts categorised in four levels: 0, 1, 2, and 3 or more. 
The results are summarised in Figure~\ref{fig:posterior_means_serious} and Figure~\ref{fig:posterior_means_slight}. 
Both graphs show a good agreement between predicted and observed number of crashes, since the distributions corresponding to higher observed counts progressively move more and more to the right. 
Moreover, Figure~\ref{fig:posterior_means_serious} shows the importance of our previous analysis and the pitfalls of predicting severe car crashes counts using the posterior means. 

\subsection{Sensitivity analysis and the modifiable areal unit problem}

Finally, we performed a sensitivity analysis evaluating the robustness of model (F) under different specifications for 1) the hyperprior distributions, 2) the adjacency matrix, and 3) the definition of the segments in the road network.

The models described in Section~\ref{sec:methodology} considered a Wishart hyperprior for the precision matrix $\bm{\Omega}$ with rank equal to $2$ and scale matrix equal to $\bm{I}_2$. 
We repeated the analysis using more vague and more informative Wishart distributions, setting the scale matrix equal to $\text{diag}(2, 2)$ and $\text{diag}(0.5, 0.5)$. 
We did not find any noticeable differences amongst alternative specifications. 
Hence results are not reported hereafter, but we refer the interested readers to the supplementary material. 

We compared different definitions for the adjacency matrix $\bm{W}$, testing second and third order neighbours and distance-based spatial neighbours, considering the first order neighbours in case two adjacent road segments are longer than the given threshold.  
Also in this case, we did not find any noticeable differences as far as the estimation of the fixed effects is concerned, whereas only small differences were found in the posterior distributions of the random effects (especially for $\sigma_{\phi_1}^2$ and $\sigma_{\phi_2}^2$ when we considered a spatial adjacency matrix with a threshold equal to $500m$). 
However, worse DIC and WAIC values were found for models using alternative definitions of $\bm{W}$ matrix, and we refer to the supplementary material for more details.  
Similar findings are also reported by \textcite{aguero2008analysis, wang2016macro, alarifi2018exploring}. 

Finally, we explored the influence of a particular configuration of the network segments on our results. 
In fact, the location of the vertices (and, hence, the edges) in a road network created with OS data is essentially arbitrary (although some minimal consistency requirements must be satisfied, see \textcite{karduni2016protocol, gilardi_lovelace_padgham_2020}), which implies that there is no unique and unambiguous way of defining the lengths and relative positions of the road segments.
We, therefore, considered an alternative network configuration reshaping and contracting the road network using an algorithm implemented in \textcite{R-dodgr}. 
This algorithm manipulates a network by excluding all \textit{redundant} vertices, i.e. those vertices that connect two contiguous segments without any other intersection \parencite{R-dodgr}.  

\begin{figure}
	\centering
	\subfloat[Road Network with redundant vertices (in red) \label{fig:redundant}]{
		\includegraphics[width=0.47\linewidth]{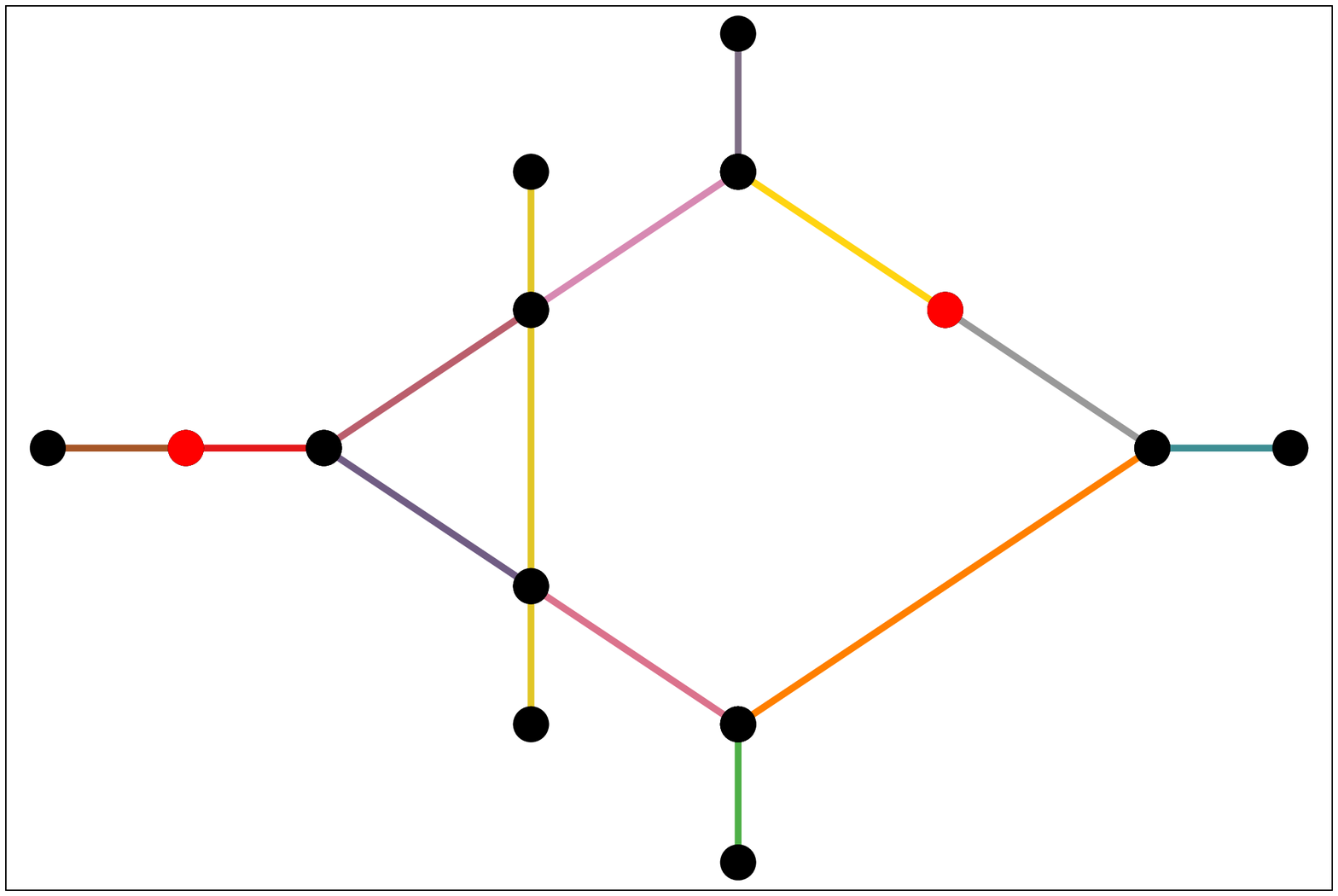}
	}
	\hspace{0.01\linewidth}
	\subfloat[Contracted road network \label{fig:contracted}]{
		\includegraphics[width=0.47\linewidth]{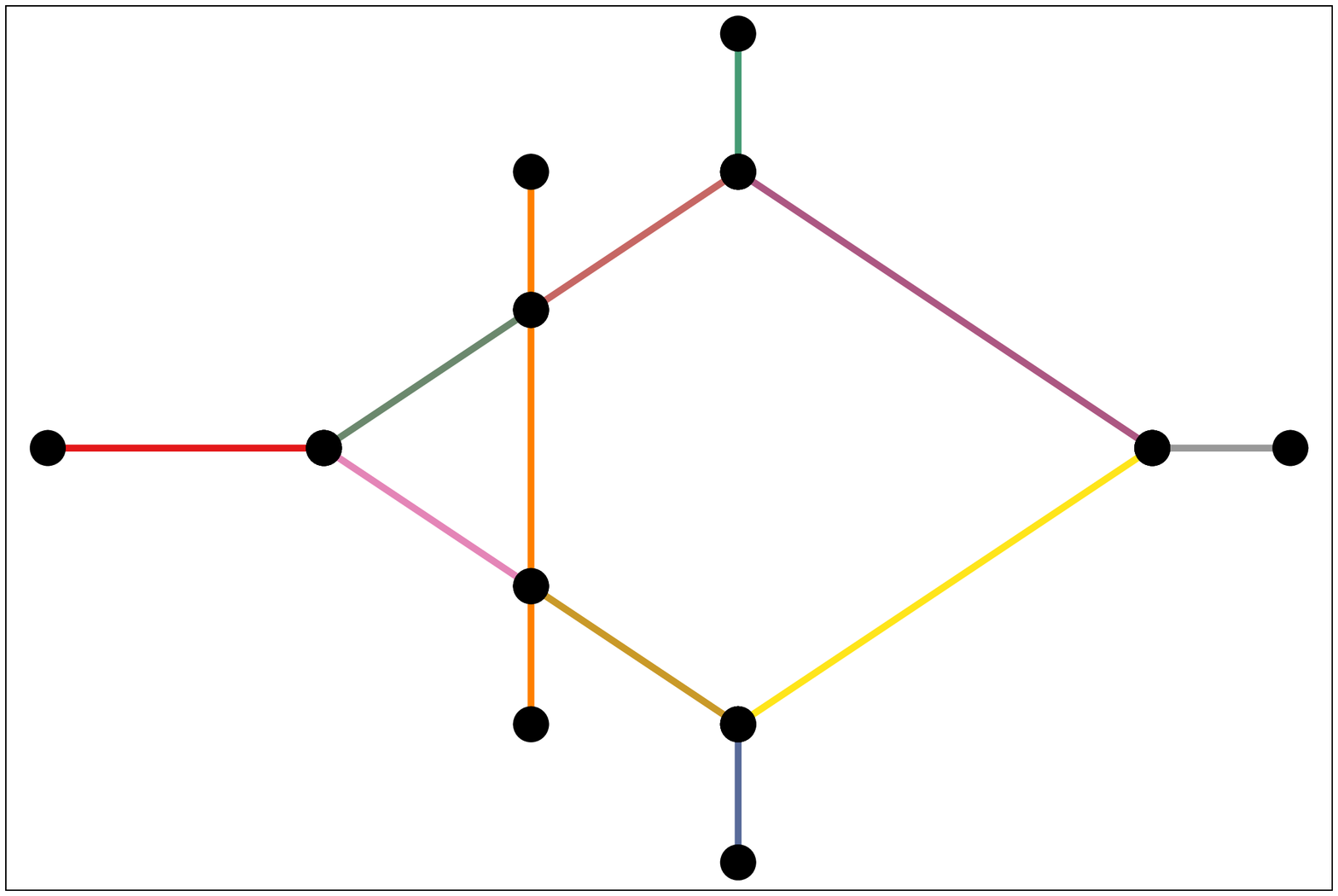}
	}
	\caption{Sketching of the algorithm used for contracting the road network. Red points on the left represent redundant vertices.}
	\label{fig:contract}
\end{figure}

A toy example representing the ideas behind the contraction of a road network is sketched in Figure~\ref{fig:contract}. 
The red dots in Figure~\ref{fig:redundant} represent redundant vertices since they can be removed without tampering the shape or the routability of the network, meaning that excluding those vertices does not add any new cluster to the graph structure.  
The goal is to remove all redundant vertices and merge the corresponding edges, creating a graph which looks identical to the original one but with fewer edges.  
Figure~\ref{fig:contracted} shows the results of the contraction operations applied to the toy network sketched in Figure~\ref{fig:redundant}. 
We can see that the redundant vertices were removed, combining the road segments that touched them.

\section{Discussion and Conclusions}
\label{sec:conclusions}

This paper investigated the spatial distribution of road crashes in a major city using Bayesian methods for road network analysis.
The relationship between crashes of different severity levels, either \textit{slight} or \textit{severe}, were modelled using a range of multivariate models to explore their spatial dynamics. 
Key to the approach was constraining crash locations to the city's road network, a one-dimensional linear network composed of segments representing a spatial lattice. 
We tested a range of multivariate hierarchical models with different random effects, and we found that the best model (according to DIC, WAIC and \textit{balanced accuracy} criteria) includes a multivariate spatially unstructured random effect and a multivariate spatially structured PMCAR random effect.  

The physical and social environment of the street segments were considered in the model specification. 
Our results, summarised in Tables~\ref{tab:fixed_effects} and~\ref{tab:random_effects}, suggest that population density is positively correlated with severe and slight crashes, although the effect is stronger for the latter case. 
On the other hand, the employment rate does not seem related with car crash occurrences, highlighting the importance of testing alternative proxies for poverty or income. 
As far as street category is concerned, Primary roads have been found safer than A roads both for severe and slight car crashes, whereas Motorways are significantly less prone to severe car accidents than A roads. 
No significant effect is found for slight accidents. 
The betweenness centrality does not seem to influence slight or severe accidents; this is an unexpected result which may deserve further investigation, possibly with alternative proxies to measure the VMT. 
Finally, dual carriageway roads have been found significantly less prone to slight car accidents, whereas no impact has been found for severe car crashes.  
Other potentially relevant physical characteristics might be the traffic speed, the line width, the presence of speed limits or junctions, the slope and the curvature of the road segments. 
Unfortunately, these variables were not available to us at the time when this paper was written.  
However, given the detailed network adopted in this paper, which is composed of thousands of short segments, we believe that some of those variables may be of limited relevance due to the short length of many segments. 
These unmeasured effects are accounted, to some extent, by the second order parameters of the random effects included in our models.

Concerning this last point, we underline that the unstructured random effects are supposed to reflect the uncertainty due to scarce sampling information (which may occur in small spatial domains like road segments) as well as the differences among the segments, such as faults in the pavements, crosses or junctions. 
These characteristics act as local shocks on the car crash occurrences and, although unrelated in space, they act both on severe and slight car accidents.
Hence, the corresponding random effects are expected to be correlated and the parameter $\rho_\theta$ is meant to measure this correlation. 

Similarly, spatially structured random effects are expected to account for the impact of unmeasured variables that have regularities is space (i.e. meteorological conditions or structural characteristics of the roads not available in our analysis). 
These variables act both on severe and slight car accidents and parameter $\rho_\phi$ is expected to measure the correlation between spatially structured random effects due to these components. 
Both correlations are found relevant in our data, although the correlation between spatially structured effects is found stronger than the other. 
Hence, we argue they both should be considered when modelling car crashes data. 
The relevant interactions between the two severity levels allow one level to borrow strength from the other and improve the estimation of the risk associated with each piece of the city road network, especially for the rarer event.

We evaluated the sensitivity of our modelling approach to different hyperprior specifications and adjacency configurations of the components of the lattice network, showing that the statistical model presents substantial robustness in this respect. 
We finally considered the impact of MAUP when modelling data collected on a spatial network. 
An algorithm was proposed in the paper to assess the magnitude of MAUP effect in the estimates and model predictions.
As mentioned above, differently from several previous studies that considered the MAUP for various areal partitions of the spatial region of interest, we found that our results are quite robust under an alternative configuration of the road network. 
This can be related to the fact that road networks have a physical meaning, hence they are expected to suffer MAUP less. 
Nevertheless, further research, possibly in different fields, is definitely necessary to better understand the impact of MAUP on network lattice data.

Finally, we remark that using areal units as spatial support ignores the fact that car crashes cannot occur outside the road network, and ascribes an estimated risk to all the streets in that polygon, whether or not they are actually exposed to it. 
Differently, using a lattice based on a network structure draws attention to limited and specific parts of the spatial support. 
Hence, adopting a more appropriate spatial disaggregation can be fundamental for local authorities to plan actions (such as to install a new traffic light, add or remove roundabouts or enforce police control) to mitigate this risk where it is found too high.

The ideas presented in this paper could be extended in several directions. 
A first step forward could be focused towards the development of a spatio-temporal extension of model (F), following the suggestions in \textcite{miaou2005bayesian, wang2011predicting, boulieri2017space, ma2017multivariate}. 
We point out, however, that this is not straightforward (and, to the best of our knowledge, it was pursued only by \textcite{ma2017multivariate} using a single road divided into a few segments) given the extreme sparse spatio-temporal nature of severe car crashes on a metropolitan road network.  
Indeed, for the dataset at hand, more than 95\% of all car crashes registered no fatal or serious car accident for any given year, something that could require a different methodological approach.
The procedure for MAUP detection could also be improved by developing new routines for testing alternative algorithms for network reshaping and contraction, which were first developed for areal data in the field of geography (see, e.g., \textcite{xu2018modifiable} and references therein). 

An additional improvement to the approach could involve the development of spatial or spatio-temporal theoretical point pattern models for car crashes on networks \parencite{baddeley2020analysing}. 
This approach represents a flexible and powerful way to investigate the spatial dynamics of random events on a graph support and may provide a tool that largely circumvents the MAUP problem. 
However, whereas it is moderately straightforward to include covariates in the bivariate lattice models suggested in this paper, the theory of bivariate point pattern models on networks and the inclusion of spatial covariates in this framework would require substantial methodological development. 
In addition, most of the work in point pattern modelling is non-parametric or semiparametric in nature, whereas the approach adopted in the present paper and, to same extent, in the analysis spatial lattice data, is grounded in the Bayesian framework. 
In this sense, the two approaches are complementary when modelling spatial or spatio-temporal dynamics of crash data. 
A bridge between the two would be the log-Gaussian Cox processes, where a stochastic component is included in the intensity function to deal with the unexplained spatial variation \parencite{Molleretal1998}. 
Dealing with an intensity function governed by a spatial random field, however, raises challenging methodological problems since defining a proper covariance function on graph spatial support is not straightforward. 
Considering this issue is beyond the scope of the present paper; hence, we do not discuss it further and leave it for future research.

It is clear that more research is needed to evaluate the full range of possible models for identifying crash `hot spots' and highlight segments on the network.
The approach presented in this paper demonstrates the potential of network-based approaches to work at city scales for flexible and robust estimates of crash rates, down to the road segment level, providing a foundation for further work in the field.

\section*{Code and data availability statement}

The data and the code that were used to estimate models from (A) to (G) and the model on the contracted network can be obtained at the following link: \url{https://github.com/agila5/multivariate-analysis-car-crashes}. 

\section*{acknowledgements}

Contains OS data © Crown copyright [and database right] 2020. 
J. Mateu is partially suported by funds PID2019-107392RB-I00 and AICO/2019/198. 
We greatly acknowledge the DEMS Data Science Lab for supporting this work by providing computational resources.

\printbibliography
\end{document}